\documentclass[a4paper,11pt]{article}
\pdfoutput=1 

\usepackage{jheppub} 

\usepackage[T1]{fontenc} 

\usepackage{amssymb} 
\usepackage{amsmath} 
\usepackage{braket} 
\usepackage{dsfont}
\usepackage{slashed} 
\usepackage{multirow} 
\usepackage{soul}
\usepackage[dvipsnames]{xcolor}

\title{\boldmath  $\eta$ and $\eta^{\prime}$ decays into lepton pairs  }

\author[a]{Pere Masjuan,}
\author[a]{Pablo Sanchez-Puertas}

\affiliation[a]{PRISMA Cluster of Excellence and Institut f\"ur Kernphysik, Johannes Gutenberg-Universit\"at, \\ Mainz D-55099, Germany}

\emailAdd{masjuan@kph.uni-mainz.de}
\emailAdd{sanchezp@kph.uni-mainz.de}

\abstract{
In this work, we calculate the branching ratios for the $\eta(\eta^{\prime})\rightarrow\overline{\ell}\ell$ decays, where $\ell = e,\mu$. These processes have tiny rates in the 
Standard Model due to spin flip, loop and electromagnetic suppressions, which could make them sensitive to New Physics effects. In order to provide a reliable input 
for the Standard Model, we exploit the general analytical properties of the amplitude. For that purpose, we invoke the machinery of Canterbury approximants, which provides a systematic description 
of the underlying hadronic physics in a data-driven fashion. Given the current experimental discrepancies, we discuss in detail the role of the resonant region and comment on the reliability of $\chi$PT calculations. Finally, we discuss the kind of new physics which we think would be relevant to account for them.
}

\begin{document}
\maketitle
\flushbottom

\section{Introduction}

Pseudoscalar decays into  lepton pairs have been subject of continuous study, starting in the fifties with the first paper by Drell~\cite{Drell:1959} on $\pi^0\rightarrow e^+e^-$. 
Special interest in these processes lies in the fact that, given the small rates which are predicted within the Standard Model (SM), they could be sensitive to New Physics 
effects~\cite{Soni:1974aw,Herczeg:1977gy,Kahn:2007ru,Chang:2008np}. The $\eta$ and $\eta^{\prime}$ cases are of particular interest since they access the $\mu^+\mu^-$ 
channel as well, opening the possibility of investigating lepton flavor violation. Given the current puzzles existing in the low-energy precision frontier of particle physics 
---specifically, the long standing discrepancy among the electron and muon anomalous magnetic moments~\cite{Agashe:2014kda,Jegerlehner:2009ry}, and the most 
recent proton radius puzzle coming from the different values obtained from electronic- and muonic-hydrogen experiments~\cite{Antognini:1900ns}, 
together with $\mathcal{R}_K$ and $\mathcal{R}_{D^{(*)}}$~\cite{Agashe:2014kda} anomalies, where $\mathcal{R}_X=\Gamma(B\to X\mu^+\mu^-)/\Gamma(B\to X e^+e^-)$---, 
it is of interest to study whether similar puzzles appear in these processes as well.

Entering this precision region requires a precise description of the underlying hadronic process fully driving this electromagnetic decay (see Fig.~\ref{fig:PtoLL}) 
as well as an accurate calculation of the loop integral involved, which itself requires a full-energy hadronic description\footnote{There is in addition a subleading $Z$-boson contribution which is necessary to be included to match our precision goal.}. Moreover, in the loop integral calculation, it has been common to neglect mass effects~\cite{Dorokhov:2008cd,Dorokhov:2009xs} which become relevant at a certain precision. In Ref.~\cite{Masjuan:2015lca}, we showed that a possible approach to avoid the inherent uncertainties from widely-used hadronic models in this calculation is provided by Canterbury approximants, which allow for a data-driven description of the hadronic process in a systematic way; this is because our method exploits the analytic properties of the underlying function. Moreover, we used an exact loop-integral numerical evaluation.
In the present work, beyond resuming the main properties of the method~\cite{Masjuan:2015lca} to  extend it to the $\eta$ and $\eta'$ cases, we shall discuss a set of novel features coming from the appearance of intermediate hadronic states absent in the $\pi^0 \to e^+e^-$, and show how our method is able to deal with them. 
Our results, together with the latest radiative corrections~\cite{Vasko:2011pi,Husek:2014tna,Husek:2015sma}, would pave the way for the precision low-energy frontier of the Standard Model and New Physics searches in these decays. 

The article is organized as follows: In Sec.~\ref{sec:intro}, we report on the state-of-the-art experimental measurements and theoretical predictions together with a general description of the main features of the process under discussion. In Sec.~\ref{sec:ca}, we review the basics of our approach based on Canterbury approximants. In Sec.~\ref{sec:toy}, we discuss the role of intermediate hadronic states and the performance of our approach with the help of a toy model. We present our results in Sec.~\ref{sec:results}. In Sec.~\ref{sec:chpt}, we discuss, in the light of our results, the role of chiral perturbation theory ($\chi$PT) when calculating such decays, together with a parameterization to obtain the $\pi^0$-exchange contribution to the $2S$ hyperfine-splitting in the muonic hydrogen; we also derive a general formula to account for the difference between the electronic and muonic channels in the SM beyond the leading order $\chi$PT estimate. Finally, with our results at hand, we discuss the implications of experimental results on New Physics searches in Sec.~\ref{sec:np}. We provide our conclusions 
and summarize our results in Sec.~\ref{ref:conclusions}.

\section{Pseudoscalar decays into lepton pairs: state of the art}
\label{sec:intro}

\subsection{Experimental status}

\begin{table}
\scriptsize
\centering
\begin{tabular}{ccccc} \hline
 $\eta\rightarrow e^+e^-$ & $\eta\rightarrow \mu^+\mu^-$ &  $\eta^{\prime}\rightarrow e^+e^-$ & $\eta^{\prime}\rightarrow \mu^+\mu^-$ &  $\pi^0\rightarrow e^+e^-$ \\ \hline
   $\leq 2.3\times10^{-6}$~\cite{Agakishiev:2013fwl} & $ 5.8(8)\times10^{-6}$~\cite{Abegg:1994wx} & $\leq 5.6\times10^{-9}$~\cite{Akhmetshin:2014hxv,Achasov:2015mek} & $-$ & $7.48(38)\times10^{-8}$~\cite{Abouzaid:2006kk} \\ \hline
\end{tabular}
\caption{Experimental results for BR$(\eta,\eta^{\prime}\rightarrow\overline{\ell}\ell)$ and  BR$(\pi^0\rightarrow e^+ e^-)$. 
}
\label{tab:exp}
\end{table} 
Pseudoscalar decays into lepton pairs are considered to be rare since their branching ratios (BR) range from $10^{-9}$ to $10^{-6}$. The state-of-the-art experimental measurements on the BR$(\eta,\eta^{\prime}\rightarrow\overline{\ell}\ell)$ with $\ell = e, \mu$ are collected in Table~\ref{tab:exp} where we have also included the $\pi^0 \to e^+e^-$ for completeness.

The $\eta\rightarrow e^+e^-$ was measured recently by the HADES collaboration in the context of dark photon searches. HADES is a fixed target experiment where a target of hydrogen or niobium is bombarded with protons and the inclusive $e^+e^-$ invariant-mass distributions are measured. In 2012, the HADES collaboration obtained with their $p+p$ data BR$(\eta \rightarrow e^+e^-) \leq 5.6 \times 10^{-6}$~\cite{HADES:2011ab}, while in 2013 they obtained, with their $p+\mathrm {Nb}$ data, BR$(\eta \rightarrow e^+e^-) \leq 2.5 \times 10^{-6}$~\cite{Agakishiev:2013fwl}. The combined outcome of the two measurements results in the PDG lower limit BR$(\eta \rightarrow e^+e^-) \leq 2.3 \times 10^{-6}$~\cite{Agakishiev:2013fwl}.

The BR$(\eta\rightarrow \mu^+\mu^-) = 5.8(8)\times10^{-6}$ quoted in the PDG~\cite{Agashe:2014kda} is a combination of the measurements performed at  SATURNE II ~\cite{Abegg:1994wx} and at Lepton-G~\cite{Dzhelyadin:1980kj}. The former was based on the $pd \to  \eta {}^{3}\mathrm {He}$ reaction, measured $114\pm 14$ events, and resulted in BR$(\eta\rightarrow \mu^+\mu^-) = 5.7(7)_{\textrm{stat.}}(5)_{\textrm{sys.}}\times10^{-6}$~\cite{Abegg:1994wx}. The later was based  on the reaction $\pi^- p \to \eta n$ with the $\eta$ reconstructed from the $\mu^+\mu^-$ invariant mass~\cite{Dzhelyadin:1980kj}, and measured BR$(\eta\rightarrow \mu^+\mu^-) = 6.5(2.1)\times10^{-6}$.

The bound BR$(\eta^{\prime}\rightarrow e^+e^-)\leq 5.6\times10^{-9}$~\cite{Akhmetshin:2014hxv,Achasov:2015mek} has been recently established after combining the upper bounds $\Gamma_{\eta^{\prime} \to e^+e^-}<0.0020$ eV and $\Gamma_{\eta^{\prime} \to e^+e^-}<0.0024$ eV measured with the SND and the CMD-3 detectors~\cite{Achasov:2015mek} at the VEPP-2000 $e^+e^-$ collider in the $e^+e^- \to \eta'$ process. The combination of these two bounds together with the total $\eta^{\prime}$ width, $\Gamma_{\eta^{\prime}} = 0.198(9)$MeV~\cite{Agashe:2014kda}, yielded the  BR$(\eta^{\prime}\rightarrow e^+e^-)\leq 5.6\times10^{-9}$. Let us note that the PDG still provides the old upper limit BR$(\eta^{\prime}\rightarrow e^+e^-)\leq 2.1 \times10^{-7}$ measured by the ND Collaboration, Novosibirsk, in 1988~\cite{Vorobev:1997wb}.

The process $\eta^{\prime} \to \mu^+ \mu^-$ is still to be measured. Even though this process could be measured using the same design as the $\eta\rightarrow \mu^+\mu^-$ done in SATURNE II, the problem now would be the large background from $\rho, \omega \to \mu^+ \mu^-$. A different set-up can be explored with the COMPASS experiment at CERN through the Primakoff effect and the high resolution of the muon pairs in the final sate\footnote{We thank Jan Friedrich for discussions along these lines.}. In addition, it has been recently suggested the possibility to measure this decay at LHCb~\cite{Huong:2016gob}. 

Finally, for completeness, we also quote the BR$(\pi^0 \to e^+e^-)=7.48(38)\times10^{-8}$~\cite{Abouzaid:2006kk}  measured by the KTeV Collaboration at FermiLab, which dominates the PDG value.

\subsection{Theoretical calculations}

The first theoretical discussion about $\eta$ decay into a lepton pair was done by Young in~\cite{Young:1967zzb}. He used a vector meson dominance (VMD) model for the parameterization of the TFF appearing in the decay. A similar but simpler VMD calculation was performed 
in~\cite{Quigg:1968zz} by Quigg and Jackson, obtaining\footnote{The original publication reports on the ratio $\Gamma_{\mu^+ \mu^-}/\Gamma_{\gamma \gamma}$ decay modes. To translate it into the BR here quoted,  we used the most recent BR$(\eta\rightarrow\gamma\gamma)=39.41\%$~\cite{Agashe:2014kda}.} BR$(\eta\rightarrow\mu^+\mu^-)=4.3\times10^{-6}$. For the $e^+e^-$ channel, they only reported the unitary bound corresponding to BR$(\eta \to e^+e^-) \geq 1.8 \times 10^{-9}$. As we will discuss later, the intuition of a unitary bound for $\eta$ and $\eta^{\prime}$ decays no longer holds due to the presence of hadronic intermediate states which reduce the final BR.

Later on, duality ideas were used by Babu and Ma to obtain\footnote{Ibid, footnote 3.} BR$(\eta\rightarrow \mu^+\mu^-)=4.89\times10^{-6}$~\cite{Babu:1982yz}.
These ideas were employed as well in quark-loop model estimations by Ametller and collaborators~\cite{Ametller:1983ec}, obtaining BR$(\eta\rightarrow\mu^+\mu^-)=1.13\times10^{-5}$~\cite{Scadron:1983nt}.  These may be refined by including a model for the $q\overline{q}$ bound state wave function~\cite{Margolis:1992gf} yielding BR$(\eta\rightarrow\mu^+\mu^-)=4.3\times10^{-6}$ and BR$(\eta\rightarrow e^+e^-)=6.3\times10^{-9}$.

The ideas from chiral perturbation theory ($\chi$PT) were implemented for the first time in these processes by Savage, Luke, and Wise in~\cite{Savage:1992ac}. The calculation of the 
loop process in $\chi$PT introduces a counterterm that should be fixed by external information. For that reason, the authors used the experimental BR$(\eta\rightarrow\mu^+\mu^-)$ 
as an input, obtaining 
BR$(\eta\rightarrow e^+ e^-)=5(1)\times10^{-9}$ as an output. This was revised later on by G\'omez Dumm and Pich in~\cite{GomezDumm:1998gw}, finding BR$(\eta\rightarrow e^+e^-)=5.8(2)\times10^{-9}$. Using SU(3) symmetry and large-$N_c$ arguments, they also quoted BR$(\eta^{\prime}\rightarrow e^+e^-)=1.5(1)\times10^{-10}$ and BR$(\eta^{\prime}\rightarrow\mu^+\mu^-)=2.1(3)\times10^{-7}$.

The development of VMD from an effective field theory point of view together with $\chi$PT led to the revision in~\cite{Ametller:1993we} by Ametller et al., which obtained\footnote{Ibid, footnote 3.} 
BR$(\eta\rightarrow\mu^+\mu^-)=4.5(^{+3}_{-1})\times10^{-6}$.

A similar study was performed by Silagadze in~\cite{Silagadze:2006rt} obtaining BR$(\eta\rightarrow \mu^+\mu^-)=5.2(1.2)\times10^{-6}$ and BR$(\eta^{\prime}\rightarrow\mu^+\mu^-)=1.4(2)\times10^{-7}$. Remarkably, this was the first calculation using an exact result for the loop integral without numerical approximations.

The first implementation of the ideas coming from the large-$N_c$ limit of QCD within the $\chi$PT framework and resonant saturation prescription was carried out by the Marseille group in~\cite{Knecht:1999gb} and obtained\footnote{Ibid, footnote 3.} BR$(\eta\rightarrow e^+e^-)=4.5(2)\times10^{-9}$ and BR$(\eta\rightarrow\mu^+\mu^-)=5.5(8)\times10^{-6}$. 
More recently, the implementation of phenomenological and theoretical constraints in models for the TFFs led the Dubna group in Ref.~\cite{Dorokhov:2007bd} to the estimates BR$(\eta\rightarrow e^+e^-)=4.60(6)\times10^{-9}$ and BR$(\eta\rightarrow \mu^+\mu^-)=5.11(20)\times10^{-6}$. Later on, the implementation of certain numerical corrections discussed in~\cite{Dorokhov:2009xs} ---neglecting errors and using a simple TFF model--- yielded in~\cite{Dorokhov:2009jd} the predictions BR$(\eta\rightarrow e^+e^-)=5.19\times10^{-9}$, BR$(\eta\rightarrow \mu^+\mu^-)=4.76\times10^{-6}$, BR$(\eta^{\prime}\rightarrow e^+e^-)=1.83\times10^{-10}$ and BR$(\eta^{\prime}\rightarrow \mu^+\mu^-)=1.24\times10^{-7}$.

The impact of the numerical corrections above show that approximate calculations do not represent a reliable result for the $\eta$ and $\eta'$ cases. Consequently, we will employ an exact numerical evaluation in what follows. For completeness, we also include the $Z$-boson contribution in our final results since its size is similar to some of the hadronic uncertainties.

\subsection{A dissection of the $P\to\overline{\ell}\ell$ process}

\begin{figure}
\centering
   \includegraphics[width=0.4\textwidth]{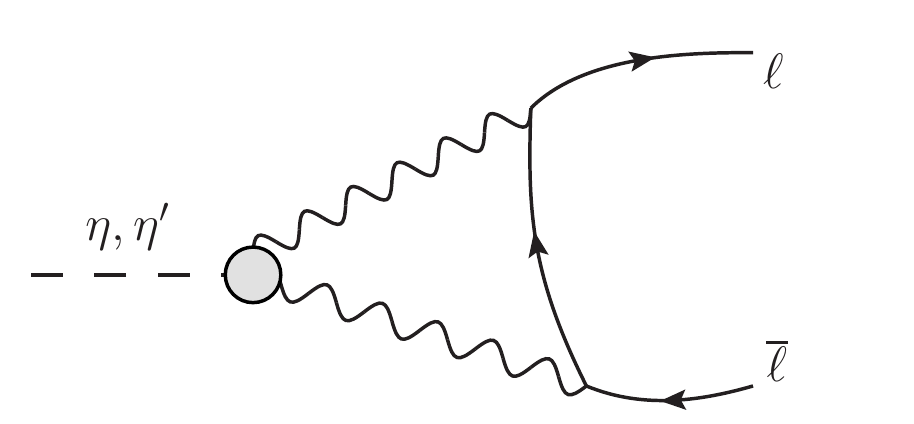}
   \caption{Leading contribution to $\eta(\eta^{\prime})$ decays into lepton pairs. The blob represents the TFF.}
   \label{fig:PtoLL}
\end{figure}

The dominant contribution to the $P\rightarrow\overline{\ell}\ell$, illustrated in Fig.~\ref{fig:PtoLL}, is mediated through an intermediate two-photon 
state, which is a loop process of the momentum $k$ running through the photons\footnote{There is an additional $Z$ boson contribution to these processes which, as shown in Sec.~\ref{sec:np}, is subleading but included in our final results.}. The main vertex of the process (the blob in  Fig.~\ref{fig:PtoLL}), describes the $P \to \gamma^* \gamma^*$ transition which is of electromagnetic nature and proceeds through the Adler-Bell-Jackiw anomaly~\cite{Adler:1969gk,Bell:1969ts}. Since the photons are virtual, the corresponding amplitude reveals the meson structure at all energies in terms of the pseudoscalar-photon transition form factors (TFFs).  The TFF, which is a function of the photon virtualities $q_1^2$ and $q_2^2$,  is denoted as $F_{P\gamma^*\gamma^*}(q_1^2,q_2^2)$. 

A description of the TFF at all energies is a formidable task because not only involves different scales but also requires knowledge of the intermediate particles produced by the photons. 
No complete description of both time-like (TL) and space-like (SL) regions is available so far. Fortunately, due to the kinematics of the process, the required information belongs 
(mostly) to the SL region, for which certain information is available. For example, $(F_{P \gamma \gamma}(0,0))^2 \sim \Gamma_{P \to \gamma \gamma}$, which are measured quantities and also calculable in the chiral limit; the large photon virtualities are known in turn from perturbative QCD~\cite{Lepage:1980fj}. 
The interpolation in between these two regimes represents still a challenge in QCD, and demands a model. Since the 
problem is how to perform the loop calculation in the SL, with scarce set of theoretical input but with great deal of experimental data with respect to the TFF at low and intermediate 
energies, our attempt in the present work is to perform an interpolation that should satisfy all possible constraints. From the mathematical point of view, this problem is called the 
\emph{general rational Hermite interpolation problem} and the solution is known to be within the mathematical theory of Pad\'e approximants~\cite{Baker}. Pad\'e approximants (PA) are rational functions $R_N(x)/Q_M(x)$ with a contact of order ${\cal O}(x^{N+M+1})$ with the function one wants to approximate. That is, their Taylor expansion matches the first $N+M+1$ terms from the original function. Only by satisfying these accuracy-through-order conditions can one claim that the analytical properties of the original function are retained. The blob in  Fig.~\ref{fig:PtoLL} represents, however, the TFF of double virtuality and the appropriate approximation to the TFF must be of a bivariate kind. The extension of PA to two variables are called Canterbury approximants~\cite{Masjuan:2015lca}.

The branching ratio can be expressed, then, in terms of the 
normalized TFF $\tilde{F}_{P\gamma^*\gamma^*}(k^2,(q-k)^2)$, where $\tilde{F}_{P\gamma^*\gamma^*}(0,0) =1$, as
\begin{equation}
\label{eq:BR}
\frac{\mathrm{BR}(P\rightarrow \overline{\ell}\ell)}{\mathrm{BR}(P\rightarrow\gamma\gamma)}  =  2 \left(\frac{\alpha_{em} m_{\ell}}{\pi m_{P}}\right)^2\beta_{\ell} |\mathcal{A}(m_{P}^2)|^2,
\end{equation}
where $\beta_{\ell} = (1-4m_{\ell}^2/m_{P}^2)^{1/2}$ is the outgoing lepton velocity and $\mathcal{A}(m_{P}^2)$ is given by the loop integral
\begin{equation}
\label{eq:loopint}
\mathcal{A}(q^2) = \frac{2i}{\pi^2q^2} \int d^4k \frac{ (q^2k^2 - (q {\color{red}{\cdot}} k)^2)\tilde{F}_{P\gamma^*\gamma^*}(k^2,(q-k)^2)}{  k^2(q-k)^2((p-k)^2-m_{\ell}^2)} .
\end{equation}
Anticipating that $|\mathcal{A}(m_P^2)|$ is typically of $\mathcal{O}(10)$, the prefactor in Eq.~\eqref{eq:BR} ---$10^{-11}(10^{-7})$ for 
$\eta\rightarrow e^+e^-(\mu^+\mu^-)$, and $10^{-12}(10^{-7})$ for $\eta^{\prime}\rightarrow e^+e^-(\mu^+\mu^-)$--- 
predicts tiny rates for these processes. 

We note that the integral Eq.~\eqref{eq:loopint} diverges logarithmically for a constant TFF, and requires, at least, a $k^{-2}$ damping at large energies, which is ensured from the OPE expansion~\cite{Novikov:1983jt}. Though this may suggest a relevant role of the large energies, it turns out that it is the low-energy region (below $1$ GeV) of the double virtual TFF which dominates in this calculation~\cite{Masjuan:2015lca}. Regretfully, such regime is poorly known and limits the applicability of data-driven approaches.

In order to perform the calculation in Eq.~\eqref{eq:loopint} ---as we have already said--- several models for the TFF have been proposed. They started with the advent of VMD ideas~\cite{Bergstrom:1983ay}. Their theoretical limitations however, arising from the use of a finite set of resonances in a narrow width approximation~\cite{Masjuan:2012sk} and lacking the connection with pQCD ---a fact circumvented with the use of Pad\'e theory~\cite{Masjuan:2007ay,Masjuan:2008fr}--- limited their reliability. To supply this, it has been proposed to use high-energy QCD constraints~\cite{Babu:1982yz} and, later, well-known phenomenological models constrained from data, which alleviate the model dependency~\cite{Dorokhov:2007bd}. 
Moreover, given the lack of double-virtual data on the TFF (which is still an experimental challenge), all the model parameters had to be fixed from high-energy QCD constraints~\cite{Dorokhov:2007bd} only. Whereas these constraints should be considered, they might not account for the low-energy behavior which, in turn, is the dominant region contributing to this process and may spoil the global description~\cite{Masjuan:2015lca}.

An alternative idea to circumvent these problems is to use $\chi$PT. We only note for the moment that this approach is not well suited though to describe the $\eta-\eta^{\prime}$ system and we will comment on its accuracy at the end of Sec.~\ref{sec:chpt}.

\section{Canterbury approximants}
\label{sec:ca}

In this work, we use the method of Canterbury approximants (CAs) to implement the TFF low-energy behavior together with the QCD constraints in a data-driven approach. 
Indeed, assuming the relevant intermediate hadronic states to be the $\rho$ resonance (through $\pi\pi$ rescattering), and the $\omega$ and the $\phi$ narrow-width resonances, all of them with positive imaginary part and belonging therefore to the class of Stieltjes functions, the convergence of CAs would be guaranteed as follows from Refs.~\cite{Chisholm,Alabiso:1974vk}. 
Given a symmetric bivariate function $F_{P\gamma^*\gamma^*}(Q_1^2,Q_2^2)=F_{P\gamma^*\gamma^*}(Q_2^2,Q_1^2)$ with known Taylor expansion
\begin{equation}
F_{P\gamma^*\gamma^*}(Q_1^2,Q_2^2) =F_{P\gamma^*\gamma^*}(0,0)\left(  1-\frac{b_P}{m_P^2}(Q_1^2+Q_2^2) + \frac{c_P}{m_P^4}(Q_1^4+Q_2^4) + \frac{a_{P;1,1}}{m_P^4}Q_1^2Q_2^2 + ... \right) ,
\end{equation}
CAs~\cite{Chisholm,Chisholm2,Masjuan:2015lca} are rational functions\footnote{We refer the interested reader to Appendix~\ref{sec:app} and to 
Ref.~\cite{Masjuan:2015lca}, where CAs were introduced.} of bivariate polynomials of degree $N$ and $M$, respectively, which coefficients are defined as to match the low-energy expansion of the original $F_{P\gamma^*\gamma^*}(Q_1^2,Q_2^2)$ function, i.e.,
\begin{equation}
C^N_M(Q_1^2,Q_2^2) \equiv \frac{R_N(Q_1^2,Q_2^2)}{Q_M(Q_1^2,Q_2^2)} \, ,
\end{equation}
fulfilling the conditions specified by Eqs.~(\ref{eq:cadeq1},~\ref{eq:cadeq2}). Only from this definition is our rational function guaranteed to converge to the original function provided it fulfills certain analytical properties (for instance, if it is a meromorphic~\cite{Baker,Baker2,Peris:2006ds,Masjuan:2007ay,Masjuan:2008fr,Masjuan:2012wy,Masjuan:2012qn} or a Stieltjes function~\cite{Baker,Baker2,Peris:2006ds,Masjuan:2008cp,Masjuan:2009wy,Alabiso:1974vk}). This construction allows to describe the TFF with the correct low-energy implementation, which is known to play the main role in these processes, a fact often overlooked (see the discussion in~\cite{Masjuan:2015lca}). Moreover, it has the ability to implement at the same time the high energy QCD-behavior, which is of relevance for the high-energy tail of the integration.

As discussed in Ref.~\cite{Masjuan:2015lca}, the simplest element of the $C^N_M(Q_1^2,Q_2^2) $ sequence reads~\cite{Masjuan:2015lca}:
\begin{equation}
\label{eq:C01}
C^0_1(Q_1^2,Q_2^2) = \frac{a_0}{1 + \frac{b_P}{m_P^2}(Q_1^2+Q_2^2) +(\frac{2b_P^2-a_{P;1,1}}{m_P^4})Q_1^2 Q_2^2},
\end{equation}
\noindent
where $a_0 = F_{P\gamma\gamma}(0,0)$ is related to the $\Gamma_{P\gamma\gamma}$, 
\begin{equation}
\left|F_{P\gamma\gamma}(0,0)\right|^{2}=
\frac{64\pi}{(4\pi\alpha)^{2}}\frac{\Gamma(P\rightarrow\gamma\gamma)}{m_{P}^{3}}\ ,
\label{Pgammagamma0}
\end{equation}
with $m_P$ the pseudoscalar mass. $F_{\eta\gamma\gamma}(0,0) = 0.274(5)$GeV$^{-1}$ and $F_{\eta^{\prime}\gamma\gamma}(0,0) = 0.344(6)$GeV$^{-1}$ using the $\Gamma_{P\gamma\gamma}$ from the PDG~\cite{Agashe:2014kda}.
The parameter $b_P$ is the slope of the single-virtual TFF~\cite{Masjuan:2012wy,Escribano:2015nra,Escribano:2015yup} and ensures the appropriate low-energy behavior up to $\mathcal{O}(Q^4)$ corrections. The most precise determinations for the slope of 
the $\eta$ and $\eta^{\prime}$ TFFs have been obtained from a data-driven procedure and read $b_{\eta} =0.576(12)$~\cite{Escribano:2015nra} and 
$b_{\eta^{\prime}} = 1.31(4)$~\cite{Escribano:2015yup}, respectively. 
In addition, for one virtuality $C^0_1(Q^2,0)$ satisfies the Brodsky-Lepage (BL)~\cite{Lepage:1980fj} high-energy $Q^2$\textit{-behavior} by construction. For the $\pi^0$, the BL asymptotic limit reads $\lim_{Q^2\to\infty}F_{P\gamma^*\gamma^*}(Q^2,0) = 2F_{\pi}Q^{-2}$ with $F_{\pi}=92.21$~MeV~\cite{Agashe:2014kda} the pion decay constant. For the $\eta$ and $\eta'$, a similar formula exists which involves both the $\eta-\eta'$ mixing parameters and the running effects of the singlet axial current~\cite{Escribano:2015nra,Escribano:2015yup}. In our case, when we state that we fulfill the correct BL $Q^2$\textit{-behavior}, we refer to the fact that our approximant satisfies $\lim_{Q^2\to\infty}Q^2F_{P\gamma^*\gamma^*}(Q^2,0) = {\cal C}\, Q^{-2}$, but without fixing the precise ${\cal C}$ coefficient (i.e. ${\cal C}=2F_{\pi}$ for the $\pi^0$ and its counterpart for the $\eta$ and $\eta'$). Not fixing ${\cal C}$ is extremely important in our method, given the relevance of the low energies  when using the simplest $C^0_1(Q_1^2,Q_2^2)$ approximant to perform the numerical evaluation of Eq.~\eqref{eq:loopint}. Matching $b_P/m_P^2$ to ${\cal C}$ instead of to the TFF's slope will spoil the TFF low-energy expansion dominant in our decays--- see discussion in Appendix~\ref{sec:app} and Table~\ref{tab:modres}.

Not only that, but the resulting description can be checked to reproduce the low-energy data on the single-virtual TFF to 
an excellent precision and the ---less relevant--- intermediate and high-energy data up to $10\%$ accuracy, see Figs.~\ref{fig:cnn1regge}~and~\ref{fig:cnn1log} in this respect as well.

Finally, $a_{P;1,1}$ is the double virtual slope of the TFF. 
Since there is no experimental data for the double virtual TFF so far, the method of fitting the experimental data with a PA sequence considered in Refs.~\cite{Masjuan:2008fv,Masjuan:2012wy,Escribano:2013kba,Escribano:2015nra,Escribano:2015yup} cannot be used to obtain the double-virtual parameters.
For this reason, we suggested in Ref.~\cite{Masjuan:2015lca} to choose a generous range for this $a_{P;1,1}$ parameter that should cover the well-known theoretical constraints at low and high energies, including the yet unknown real value. It turns out that, at low energies 
---the most important region in our calculation--- $\chi$PT calculations favor the so called factorization approach~\cite{Bijnens:2012hf}, namely, that 
$\tilde{F}_{P\gamma^*\gamma^*}(Q_1^2,Q_2^2) = \tilde{F}_{P\gamma^*\gamma}(Q_1^2,0) \times \tilde{F}_{P\gamma\gamma^*}(0,Q_2^2)$, implying that $a_{P;1,1}=b_{P}^2$ 
in Eq.~\eqref{eq:C01}\footnote{Actually, the recent results from dispersive analysis~\cite{Xiao:2015uva} seem to confirm the factorization behavior employed in~\cite{Masjuan:2015lca} for the $\eta$ case. However, as we will show later, this approach is only safe at very low energies and the region from $1-2$~GeV$^2$ is already dominated by the pQCD behavior.}. 
On the other hand, it is known that at high energies the operator product expansion (OPE) dictates that $F_{P\gamma^*\gamma^*}(Q^2,Q^2)\sim Q^{-2}$~\cite{Babu:1982yz,Dorokhov:2007bd}, implying $a_{P;1,1}=2b_P^2$ in Eq.~\eqref{eq:C01}\footnote{Again, here we only fulfill the $Q^2${\textit{-behavior}}, i.e., that $\lim_{Q^2\to\infty}F_{P\gamma^*\gamma^*}(Q^2,Q^2) \propto Q^{-2}$, but we do not impose the particular coefficient, which turns out to be $1/3$ of the BL behavior~\cite{Babu:1982yz,Dorokhov:2007bd}.}. Deviations from factorization should translate into $a_{P;1,1}>b_P^2$. Then, from now on and due to the ignorance on the precise value of $a_{P;1,1}$, we will assume for the rest of our work $b_P^2 \leq a_{P;1,1} \leq 2b_P^2$. In addition, we note that from Eq.~\eqref{eq:C01}, $a_{P;1,1} \leq 2b_P^2$ is required for avoiding poles in the SL region.

To scrutinize a bit more the interplay of these two regimes, let us look briefly at the $\pi^0\rightarrow e^+e^-$ decay, for which is possible to employ a reliable approximated version for Eq.~\eqref{eq:loopint}~\cite{Masjuan:2015lca} thanks to the smallness of the $\pi^0$ mass. This allows to split the integral $(I)$ in Eq.~\eqref{eq:loopint} in two parts as $I = \int_0^{\Lambda} \textrm{Fact} +\int_{\Lambda}^{\infty} \textrm{OPE}$, in which the low-energy part below the scale $\Lambda$ is integrated using the factorization approach (i.e., $a_{P;1,1}=b_{P}^2$), whereas the high-energy part above $\Lambda$ is integrated using the OPE ($a_{P;1,1}=2b_{P}^2$). We obtain BR$(\pi^0\rightarrow e^+e^-)=\{6.15,6.21,6.26,6.29\}\times10^{-8}$ for $\Lambda=\{0,1,2,\infty\}$. This shows that, if factorization represents a good approximation at low energies, the freedom on choosing when the OPE should contribute implies a non-negligible systematic error at the precision we are aiming ---of the order of the percent level--- and vice-versa. It is reasonable to assume then, that such deviation should be covered by choosing a conservative range $a_{P;1,1}\in(b_P^2,2b_P^2)$.
In addition, we remind that the mathematical theory of CAs does not dictate what  $a_{P;1,1}$ to take and both choices are equally correct as they only correspond to different reconstructions. We illustrate these discussions, e.g. 
the relevant range for the $a_{P;1,1}$ parameter and the relevance of using the low-energy parameters, in the Table~\ref{tab:modres} of Appendix~\ref{sec:app} with the help of two well motivated toy models. 
Again, we emphasize that the reconstructed element, $C^0_1(Q_1^2,Q_2^2)$, fulfill the different $Q^2$-{\textit{behaviors}} which are implied by QCD for $a_{P;1,1}=2b_P^2$, but not the corresponding coefficients  
---these could be implemented in higher elements, which unfortunately involve at the moment too many unknown parameters of double virtuality.
Still, from our studies in~\cite{Masjuan:2015lca} together with Appendix~\ref{sec:app} and the reasons presented above, these effects are seen to be tiny and the quoted value already provides a reliable result.

The last ingredient in our approach
is the estimation of a systematic error induced by the truncation of the CA sequence, i.e., the fact that even though the CA sequence converges to the TFF, at a given and finite CA order, the difference between the function and its approximant is not zero. The inclusion of a systematic error marks a difference with respect to previous studies. This is the goal of the next chapter.

\section{Assessing the systematic error}
\label{sec:toy}

\begin{figure}
\centering
   \includegraphics[width=0.8\textwidth]{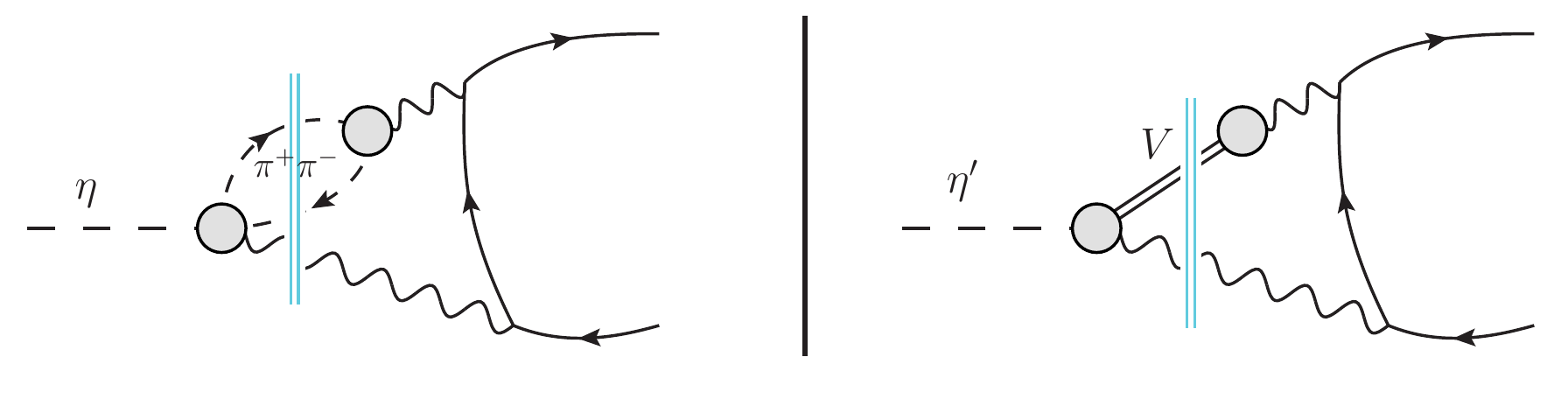}
   \caption{Intermediate hadronic states invalidating the unitary bound for the $\eta(\eta^{\prime})$, left(right).}
   \label{fig:UB}
\end{figure}

The $\eta$ and $\eta^{\prime}$ masses are large enough to yield intermediate hadronic states in the $P \to \bar{\ell} \ell$ processes as sketched in Fig.~\ref{fig:UB}, which implies an additional imaginary part beyond that of the $\gamma\gamma$ contribution via Cutcosky rules. As we have already said, this novel feature ---not present for the $\pi^0$--- diminishes the imaginary part with respect to the $\gamma\gamma$ intermediate state invalidating the unitary bound~\cite{Drell:1959}, $|\mathcal{A}(m_P^2)|^2 \geq \textrm{Im}(\mathcal{A}_{\gamma\gamma}(m_P^2))^2 $, which tacitly assumes the absence of intermediate states beyond the $\gamma\gamma$ one. 
This effect can be understood from the loop integral as well. After Wick-rotation, the TFF integration domain lies in the $-m_P^2\leq Q^2 \leq \infty$ region, which for the 
$\eta$ and $\eta^{\prime}$ involves threshold production ($m_{\eta}>2 m_{\pi}$) and resonances ($m_{\eta^{\prime}} > m_{\rho,\omega}>2 m_{\pi}$). 

This effect has never been considered before when calculating the rare decays and must be taken into account as well when evaluating the systematic error. As we will show below, the uncertainty from the resonance part of Fig.~\ref{fig:UB} becomes the dominant source of error. To quantitatively study this effect, we take a toy model for the TFF that includes both a two-pion production threshold and a vector resonance. The model is conceived in such a way that the time-like region contains all the required features of the physical TFF up to the $\eta^{\prime}$ mass. 
The first ingredient in our toy model is factorization, which as explained before seems a reasonable choice at low energies. The second ingredient is the use of vector meson dominance ideas~\cite{Landsberg:1986fd} allowing to express the single-virtual TFF as 
\begin{equation}
\label{eq:VMD}
\tilde{F}_{P\gamma^*\gamma}(s) = c_{P\rho} G_{\rho}(s) + c_{P\omega} G_{\omega}(s) + c_{P\phi} G_{\phi}(s),
\end{equation}
where $G_V(s)$ are the different resonance contributions weighted by the dimensionless couplings $c_{PV}$ which are obtained from a quark-model, 
$c_{\eta(\eta^{\prime})\rho}=9/8(9/14)$, $c_{\eta(\eta^{\prime})\omega}=1/8(1/14)$, $c_{\eta(\eta^{\prime})\phi}=-2/8(4/14)$~\cite{Hanhart:2013vba}, and $G_V(0)=1$. 
In order to incorporate the $\pi\pi$ intermediate branch cut from Fig.~\ref{fig:UB}, fulfilling unitarity and analyticity, we take for the $\rho$ contribution $G_{\rho}(s)$ a model based on Refs.~\cite{GomezDumm:2000fz,Dumm:2013zh}
\begin{equation}
\label{eq:rho}
G_{\rho}(s) = \frac{M_{\rho}^2}{   M_{\rho}^2 - s +\frac{s M_{\rho}^2}{96\pi^2F_{\pi}^2}\left( \ln\left(\frac{m_{\pi}^2}{\mu^2}\right)  +\frac{8m_{\pi}^2}{s} -\frac{5}{3} - \sigma(s)^3 \ln\left(\frac{\sigma(s)-1}{\sigma(s)+1}\right)   \right)   }
\end{equation}
with $\sigma(s) = \sqrt{1-4m_{\pi}^2/s}$, and the parameters $M_{\rho}=0.815$ GeV, $F_{\pi}=0.115$ GeV, $\mu=0.775$ GeV, and $m_{\pi}=0.139$ GeV, chosen to reproduce the pole 
position $s_{\rho}=(M - i \Gamma/2)^2$ with $M =  0.764$ GeV and $\Gamma = 0.144$ GeV from~\cite{Masjuan:2013jha}, while for the (narrow-width) $\omega,\phi$ resonances, we take\footnote{We explored further
refined models with an improved threshold behavior for the $\omega$ and $\phi$ resonances. Given their narrow width they led to very similar results and we decided to take the ones in Eq.~\eqref{eq:omegaphi} for not obscuring our study and deviating the attention from our main concern, an estimation of a systematic error.}
\begin{equation}
\label{eq:omegaphi}
 G_{\omega,\phi} = \frac{M_{\omega,\phi}^2 + M_{\omega,\phi}\Gamma_{\omega,\phi}(s_{th}/M^2_{\omega,\phi})^{3/2}}{M_{\omega,\phi}^2 - s + M_{\omega,\phi}\Gamma_{\omega,\phi} ((s_{th}-s)/M^2_{\omega,\phi})^{3/2}},
\end{equation}
which parameters are fixed from PDG masses and widths~\cite{Agashe:2014kda}. 
This choice makes our model very similar to the dispersive approach formulated in~\cite{Hanhart:2013vba}. 

To evaluate now the branching ratio of the decay, we have to calculate the diagram in Fig.~\ref{fig:PtoLL}. The blob there stands for the TFF, which includes among others the contributions from Fig.~\ref{fig:UB} and is accounted for by the model in Eq.~(\ref{eq:VMD}). Employing the Cauchy integral representation for the (factorized) TFF,
\begin{equation}
\label{eq:cauchy}
F_{P\gamma^*\gamma^*}(q_1^2,q_2^2) = \frac{1}{\pi^2} \int_{s_{th}}^{\infty} dM_1^2  \int_{s_{th}}^{\infty} dM_2^2  \frac{\textrm{Im}[F_{P\gamma^*\gamma}(M_1^2)]}{q_1^2 - M_1^2 -i\epsilon} \frac{\textrm{Im}[F_{P\gamma^*\gamma}(M_2^2)]}{q_2^2 - M_2^2 -i\epsilon} , 
\end{equation}
and changing the integration order in the integral \eqref{eq:loopint}, allows to express the loop amplitude as 
\begin{align}
\label{eq:loopdisp}
\mathcal{A}(q^2)   = & \frac{1}{\pi^2}  \int_{s_{th}}^{\infty} dM_1^2  \int_{s_{th}}^{\infty} dM_2^2  \textrm{Im}[\tilde{F}_{P\gamma^*\gamma}(M_1^2)]  \textrm{Im}[\tilde{F}_{P\gamma^*\gamma}(M_2^2)] \nonumber \\
 &  \times   \left(\frac{2i}{\pi^2q^2} \int d^4k \frac{ (q^2k^2 - (q k)^2)}{ k^2(q-k)^2((p-k)^2-m_{\ell}^2)}\frac{1}{k^2-M_1^2}\frac{1}{(q-k)^2-M_2^2}\right) \nonumber \\
 \equiv   &   \frac{2}{\pi^2} \int_{s_{th}}^{\infty} dM_1^2  \int_{s_{th}}^{M_1^2} dM_2^2\,  \textrm{Im}[\tilde{F}_{P\gamma^*\gamma}(M_1^2)]  \textrm{Im}[\tilde{F}_{P\gamma^*\gamma}(M_2^2)]  \times      K(M_1^2,M_2^2).
\end{align}
This procedure results on an easy evaluation of the loop amplitude denoted as $K(M_1^2,M_2^2)$ through standard one-loop techniques~\cite{'tHooft:1978xw} or a numerical evaluation using \texttt{LoopTools}~\cite{Hahn:1998yk}. 

\begin{figure}
\centering
\includegraphics[width=0.6\textwidth]{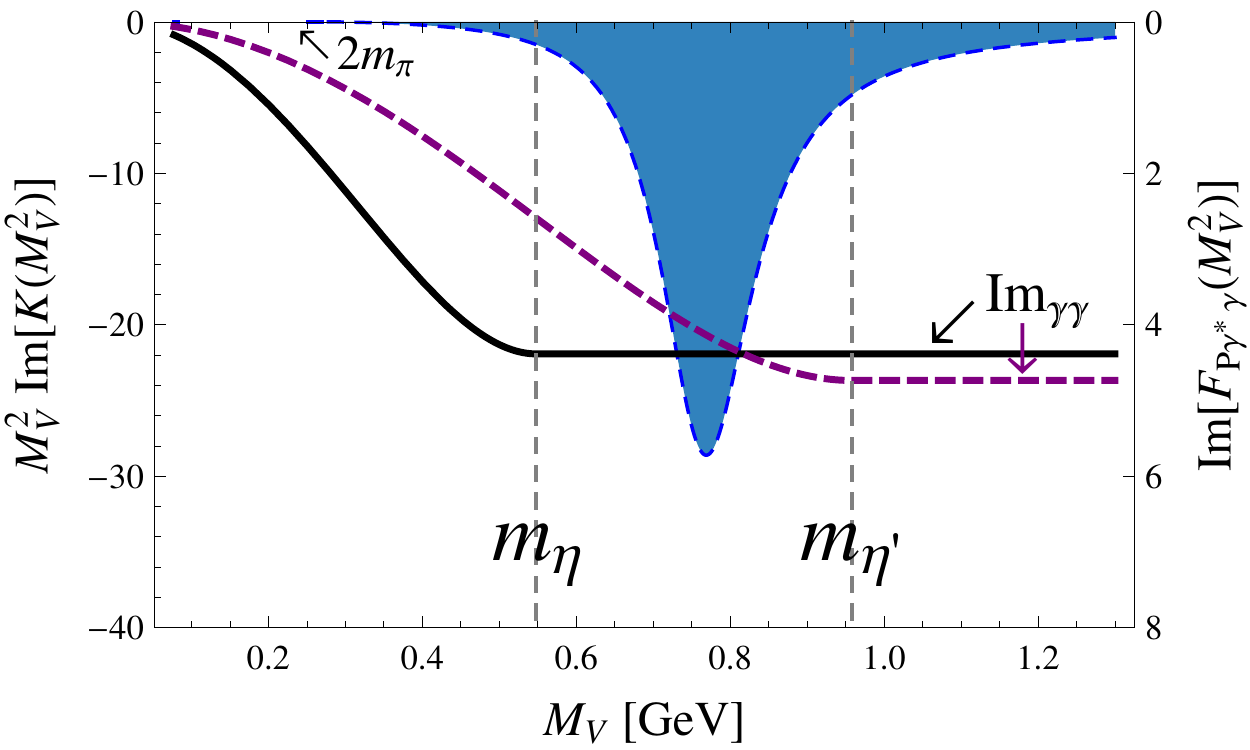}
\caption{The imaginary part for integrand \eqref{eq:loopdisp} expressed in terms of $M_V^2\textrm{Im}[K(M_V^2)]$ (black and dashed-purple lines for the 
$\eta$ and $\eta'$, respectively) which has then to be convoluted with $\textrm{Im}[\tilde{F}_{P\gamma^*\gamma}(M_V^2)]$. As an illustration, we show 
$\textrm{Im}[\tilde{F}_{P\gamma^*\gamma}(M_V^2)] = \textrm{Im}[G_{\rho}(M_V^2)]$ alone as a bluish area ---the $\textrm{Im}[G_{\omega,\phi}(M_V^2)]$ contributions would 
appear as sharp peaks at $M_V=m_{\omega,\phi}$.}
\label{fig:imdisp}
\end{figure}

Now, the threshold effects are clear and easier to handle. To illustrate them, we plot the imaginary part of the integrand in \eqref{eq:loopdisp} in terms of 
$\textrm{Im}[\tilde{F}_{P\gamma^*\gamma}(M_V^2)]$ and $\textrm{Im}[K(M_V^2)]$ ---which contains both $\gamma \gamma$ and vector contributions--- when dispersing only one virtuality in~\eqref{eq:loopdisp} for simplicity (i.e., we neglect the $q^2$ dependence on the second virtuality). The resulting plot is shown in Fig.~\ref{fig:imdisp} as a solid-black (dashed-purple) line for the $\eta (\eta^{\prime})$ in terms of the dispersive variable $M_V$ once the $\int d^4k$ integration has been performed to give $K(M_V^2)$ in the last line of Eq.~\eqref{eq:loopdisp}. These lines have to be convoluted with $\textrm{Im}[\tilde{F}_{P\gamma^*\gamma}(M_V^2)]$. In Fig.~\ref{fig:imdisp}, this is represented by the bluish area. For clarity, we only plot there $\textrm{Im}[G_{\rho}(s)]$, since the $\rho$ resonance is the only one relevant in discussing $\pi\pi$ threshold effects. The overall $\textrm{Im}[\tilde{F}_{P\gamma^*\gamma}(M_V^2)]$ contribution would resemble thereby that in Fig.~\ref{fig:imdisp}, but with additional sharp peaks at $M_V=m_{\omega},m_{\phi}$ locations, each of them weighted by the corresponding $c_{\eta(\eta')V}$ coefficient. For $M_V>m_P$, the $\gamma\gamma$ contribution dominates (the bluish region below $m_{\eta}$ in Fig.~\ref{fig:imdisp} is almost negligible) and will be slightly modified when $m_P > 2 m_\pi$ due to the tail of the resonance contribution, whereas it will be less important when $M_V<m_P$. Then, whenever $\textrm{Im}[F_{P\gamma^*\gamma}(M_V^2< m_P^2)]\neq0$ , $\textrm{Im}[\mathcal{A}(m_P^2)]$ will be shifted with respect to the $\gamma \gamma$ contribution. This is, as unitarity implies, whenever an intermediate hadronic channel appears below $m_P$.

For completeness, we illustrate in Table~\ref{tab:im} the numerical shift in the imaginary part induced by the vector contributions with respect to the 
$\gamma\gamma$ one using our toy model~\eqref{eq:VMD}---we stress that the three $(V=\rho,\omega,\phi)$ channels have been included in Table~\ref{tab:im} 
calculations. The resulting shift illustrates the break of the unitary bound.
\begin{table}[h]
\centering
\begin{tabular}{cccc}\hline
 &  & $\gamma\gamma$ & Total \\ \hline
\multirow{2}{*}{$\textrm{Im}[\mathcal{A}_{\eta\rightarrow \ell\ell}(m_{\eta}^2)]$}    & $ee$         & $-21.920$ & $-21.805$ \\ 
                                                                                                                   & $\mu\mu$ & $-5.468$ & $-5.441$ \\ \hline
\multirow{2}{*}{$\textrm{Im}[\mathcal{A}_{\eta^{\prime}\rightarrow \ell\ell}(m_{\eta^{\prime}}^2)]$}   & $ee$         & $-23.675$ & $-19.251$ \\ 
                                                                                                                   & $\mu\mu$ & $-7.060$ & $-5.733$ \\ \hline
\end{tabular}
\caption{Imaginary part of $\mathcal{A}(q^2)$ (total) compared to the imaginary part calculated from the $\gamma\gamma$ channel alone. The hadronic contributions lower the total value of the imaginary part with respect to the $\gamma\gamma$ contribution, invalidating the unitary bound.}
\label{tab:im}
\end{table}

Given that our model is a Stieltjes function, it is well known that the $C^N_{N+1}(Q_1^2,Q_2^2)$ sequence is guaranteed to converge in the whole complex plane, except along the cut~\cite{Baker,Baker2}, where zeros and poles of our CA will clutter to reproduce the discontinuity~\cite{Baker,Baker2,Masjuan:2009wy}\footnote{The reader may notice that, for the $\eta$ case, the narrow $\phi$ contribution produces a shift in sign near $m_{\phi}$ for the imaginary part. We remark that in the limit of an infinitely narrow resonance, the resulting function is a combination of a Stieltjes and a meromorphic function, for which convergence applies as well~\cite{Baker}. Even if the particle would have a finite width, its effect would not affect  convergence before the sign actually changes, which poses no problem for the $\eta$ case.}. 
Remarkably, even if the sequence does not converge along the cut, the integral along its imaginary part converges globally to that of the real function, which follows 
from Cauchy's integral theorem. This means in practice that the approximant's poles
will be responsible for effectively generating an imaginary part in our integral via the $i\epsilon$ prescription mimicking the cut contribution. As an illustration, we collect the results for both BR and $\mathcal{A}(m_P^2)$ from our simplest approximant, the $C^0_1(Q_1^2,Q_2^2)$, in Table~\ref{tab:c01res} and compare its results with the toy model. The comparison of the BRs reveals a systematic error induced by the fact that we have truncated the CA sequence. For the $\eta$, such error is almost negligible (the role of the vector resonances is very mild there), whereas for the $\eta^{\prime}$ it goes almost up to $20\%$. These percentages will be used as an estimate of our systematic error in our final results for the $C^0_1(Q_1^2,Q_2^2)$ element.

We would like to remark at this point that using a VMD model with the $\rho$ mass ---which was standard in the past for performing this calculation---, we would have found BR$(\eta\rightarrow e e) = 5.30\times10^{-9}$, which implies a larger systematic uncertainty compared to our results with the $C^0_1(Q_1^2,Q_2^2)$ collected in Table~\ref{tab:c01res}, i.e., BR$(\eta\rightarrow e e) = 5.42\times10^{-9}$. Using experimental data from the space-like region to fit the VMD model does not improve on the result. In this case, we would have obtained BR$(\eta\rightarrow e e) =5.26\times10^{-9}$. 
These results illustrate the potential large systematic error coming from the usage of VMD data-fitting procedures from high energies for processes which are low-energy dominated, even if the quality of the fit is good enough and the errors tiny.\\
\begin{table}
\centering
\scriptsize
\begin{tabular}{@{\extracolsep{4pt}} ccccccc} 
\cline{1-4}\cline{5-7}	
 $BR(P\rightarrow\ell\ell)$ & toy model & $C^0_1$ & Error ($\%$) & $\mathcal{A}(m_P^2)$ & toy model & $C^0_1$\\ \cline{1-4}\cline{5-7}	
  $(\eta\rightarrow e e)\times10^{-9}$ & $5.4095$ & $5.4179$ & $0.16$ &  $(\eta\rightarrow e e)$ & $31.4-21.8i$ & $31.4-21.9i$\\
  $(\eta\rightarrow \mu\mu)\times10^{-6}$ & $4.49361$ & $4.52701$ & $0.74$ &   $(\eta\rightarrow \mu\mu)$ & $-1.09-5.44i$ & $-1.05-5.47i$\\  \cline{1-4}\cline{5-7}
  $(\eta^{\prime}\rightarrow e e)\times10^{-10}$ & $1.70507$ & $1.88331$ & $9$ &  $(\eta^{\prime}\rightarrow e e)$ & $46.4-19.2i$ & $48.7-20.5i$\\
  $(\eta^{\prime}\rightarrow \mu\mu)\times10^{-7}$ & $1.1953$ & $1.46089$ & $18$ &   $(\eta^{\prime}\rightarrow \mu\mu)$ & $3.09-5.73i$ & $3.82-6.10i$\\ 
  \cline{1-4}\cline{5-7}	
\end{tabular}
\caption{Comparison between our toy model result and the simplest $C^0_1(Q_1^2,Q_2^2)$ approximation for each channel. The Error represents the relative deviation between the model and the approximation. Left table collects the BR, whereas the right table contains the loop amplitude $\mathcal{A}(m_P^2)$. See details in the main text.}
\label{tab:c01res}
\end{table} 

\begin{figure}
\centering
   \begin{minipage}[b]{0.49\textwidth}
       \centering
       \includegraphics[width=\textwidth]{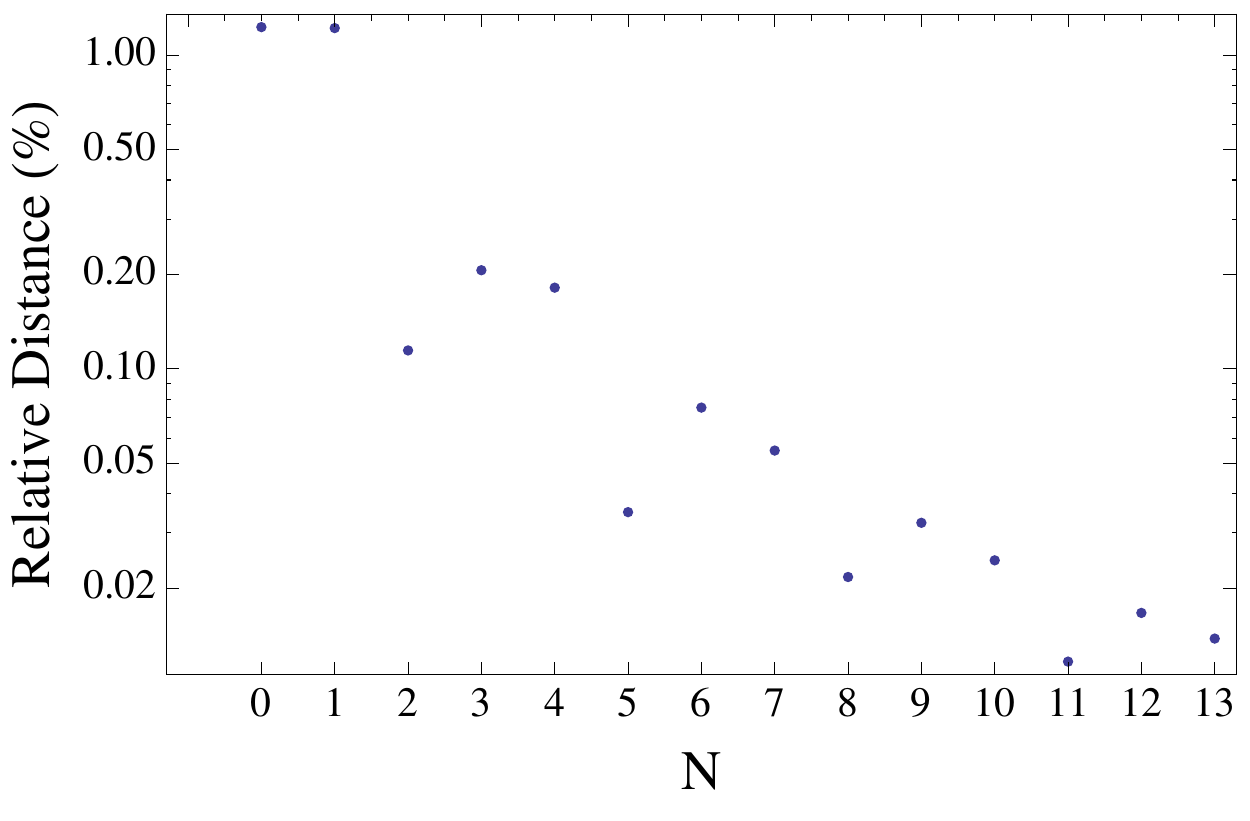}   
   \end{minipage} 
      \hspace{0.04\textwidth}
   \begin{minipage}[b][4.2cm][c]{0.45\textwidth}
       \caption{Top (bottom left) panel represent the $C^N_{N+1}$ pattern of convergence to $|\mathcal{A}(m_P^2)|$ for the toy model Eq.~(\ref{eq:VMD}) for $\eta(\eta^{\prime})$. Bottom right figure represents the poles (open circles) and zeros (slashes) from our approximants. $N$ stands for the $C^N_{N+1}$ index.  \label{fig:pacut}}
   \end{minipage}
   \includegraphics[width=0.49\textwidth]{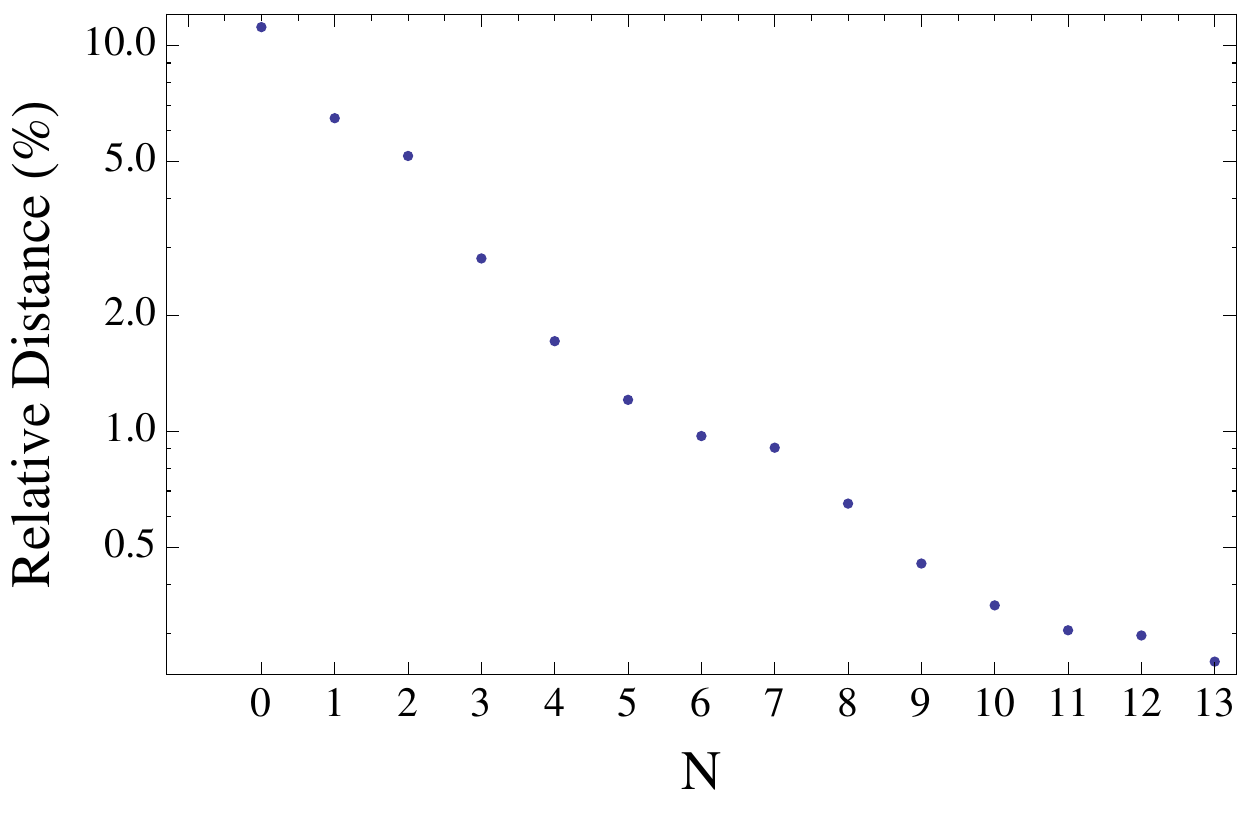}
   \includegraphics[width=0.49\textwidth]{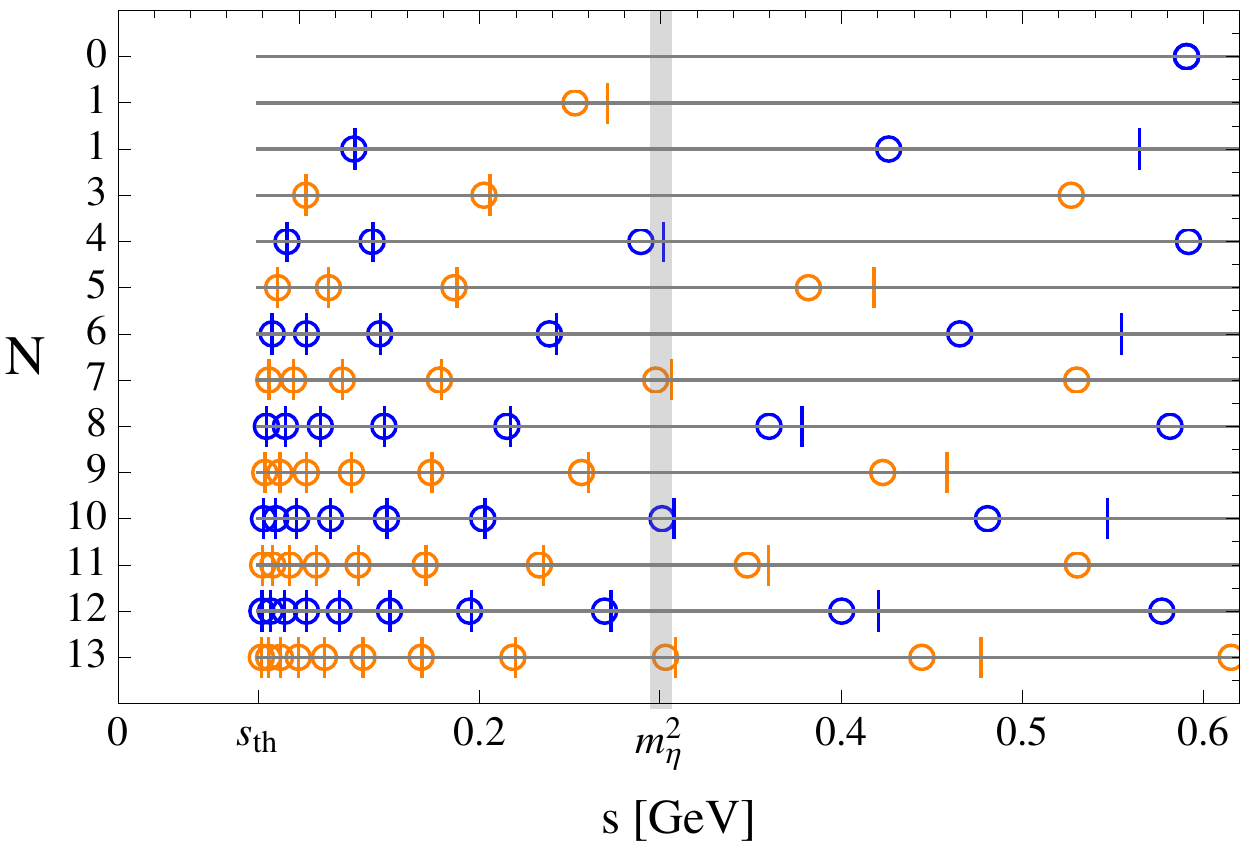}
\end{figure}

As we have said, the convergence of the CA sequence to our toy model is guaranteed in advance~\cite{Baker,Baker2,Masjuan:2009wy}. To fully show this feature beyond the simplest $C^0_1(Q_1^2,Q_2^2)$ discussed in Table~\ref{tab:c01res}, we now discuss the the
the $C^N_{N+1}(Q_1^2,Q_2^2)$ sequence up to $N=13$ and calculate the corresponding amplitude in Eq.~\eqref{eq:loopint}. 
The relative distance in the complex plane of the energy squared, defined as $|1-\mathcal{A}(m_P^2)^{\textrm{CA}}/\mathcal{A}(m_P^2)|$, is shown in Fig.~\ref{fig:pacut}, where, for simplicity, we employ only the $G_{\rho}$ contribution in~\eqref{eq:VMD} (without $\omega, \phi$ contributions).

The results in  Fig.~\ref{fig:pacut} show the ability of our approximants to systematically account for the TFF to arbitrary precision since the relative distance decreases when the order of the CA increases, even in the presence of  the non-trivial behavior of the branch cut from the intermediate hadronic states. This exercise is a proof of concept that even though the CAs cannot reproduce the imaginary part locally, they are able to approximate it globally with great precision. Note the \textit{a priori} irregular convergence for the $\eta$ case in Fig.~\ref{fig:pacut} (top panel). This is just an accident due to the combination of the particular TFF we are using together with the particular value of the $\eta$ mass. Whenever some pole is located close to the $\eta$ mass, it leads to a bad determination. This is compensated in higher approximants with a nearby zero to this pole, alleviating this effect and making it negligible as $N\rightarrow\infty$ as shown in Fig.~\ref{fig:pacut} (bottom right panel), where the poles and zeros for different approximants are plotted. 

We find then that, for the $C^N_{N+1}(Q_1^2,Q_2^2)$ element, the systematic error can be accounted for, essentially, by the difference in the BR with respect to the $C^{N-1}_{N}(Q_1^2,Q_2^2)$ result. As in our case of study we only reach the $C^0_1(Q_1^2,Q_2^2)$ approximant, this procedure does not apply and we take as the systematic error for the BR the one which is displayed in the fourth column in Table~\ref{tab:c01res} (based in our toy model). This possibly overestimates the systematic error, see comments in Sec.~\ref{sec:results}, but we opt for this to remain on the conservative side. As soon as experimental data on the double virtual TFF will be available, we will be able to extend our CA sequence and reduce the systematic error.

In Ref.~\cite{Masjuan:2015lca} we considered a similar method for establishing the systematic error in the case of the $\pi^0 \to e^+ e^-$ case. There, we used the large-$N_c$ Regge model from Ref.~\cite{RuizArriola:2006ii}  ---which roughly factorizes at low energies, obeys the OPE, and for which $a_{P;1,1}\gtrsim b_P^2$--- and obtained rather small $(\mathcal{O}(1\%))$ systematic errors for the $C^0_1(Q_1^2,Q_2^2)$ element, which was accounted for by the chosen $a_{P;1,1}$ range. As remarked before, the large systematic error in Table~\ref{tab:c01res} comes form the presence of the resonance region and the difficulty of approximating it with a single pole.

\section{Final results}
\label{sec:results}

Our final results for the pseudoscalar decays into lepton pairs using the method here described have been calculated using Eq.~(\ref{eq:loopint}) filled with the form factor data-driven parameterization described by Eq.~(\ref{eq:C01}). 

The loop integral evaluation has been performed using {\texttt{FeynCalc}}~\cite{Mertig:1990an} to obtain the decomposition into the Passarino-Veltman functions; the latter 
have been numerically evaluated using \texttt{LoopTools}~\cite{Hahn:1998yk}. In the factorization case ($a_{P;1,1}=b_P^2$) this is possible using partial fraction decomposition, 
allowing to express the TFF and photon propagators in Eq.~(\ref{eq:loopint}) as
\begin{equation}
\frac{\tilde{F}_{P\gamma^*\gamma^*}(k^2,(q-k)^2)}{k^2(q-k)^2} = \left(  \frac{1}{k^2 - \frac{m_P^2}{b_P}}  - \frac{1}{k^2}  \right)  \left(  \frac{1}{(q-k)^2 - \frac{m_P^2}{b_P}}  - \frac{1}{(q-k)^2}  \right), 
\end{equation}
and the corresponding result involves one-, two- and three-point functions. For the OPE choice ($a_{P;1,1}=2b_P^2$) the loop integration is possible rewriting the TFF as 
\begin{equation}
   \tilde{F}_{P\gamma^*\gamma^*}(k^2,(q-k)^2) = \frac{-m_P^2/2b_P}{(k-\frac{q}{2})^2 - (\frac{m_P^2}{2b_P} -\frac{q^2}{4})},
\end{equation}
which allows to express the loop integral in terms of one-, two-, three- and four-point functions.

Later on, the branching ratios have been calculated with the formula given in Eq.~(\ref{eq:BR}). The results for the amplitude $\mathcal{A}(m_P^2)$ and for the BRs are collected separately for illustrative purposes in Table~\ref{tab:mainamp}, second column, and in Table~\ref{tab:mainres}, second column,  respectively, and represent the main result of this work. 
In addition, we have included for completeness the $Z$-boson contribution (third column in Table~\ref{tab:mainamp})---we refer the interested reader to Section~\ref{sec:np} or to Ref.~\cite{Arnellos:1981bk}. Finally, these tables also include results when the kernel in Eq.~(\ref{eq:loopint}) is expanded in terms of $m_l/m_P$ as well as $m_l/\Lambda$ and $m_P/\Lambda$, with $\Lambda$ a cut-off of the loop integral (fourth column in Tables~\ref{tab:mainamp} and \ref{tab:mainres}).

To report a final result, several sources of errors must be taken into account:
\begin{itemize}
\item Our ignorance on the value of the slope of double virtuality $a_{P;11}$, Eq.~(\ref{eq:C01}), is captured by the range $a_{P;11}\in(b_P^2\div 2b_P^2)$. With this range, we obtain the results for the loop amplitude $\mathcal{A}(m_P^2)$, Eq.~\eqref{eq:loopint}, collected in Table~\ref{tab:mainamp}, and its associated BR displayed in Table~\ref{tab:mainres}. The values, which are reported in a range form corresponding to imposing the OPE or the factorization, include statistical errors for the amplitudes and statistical and systematic errors for the branching ratios\footnote{The systematic error shown in Table~\ref{tab:c01res} (column called Error) and discussed in Sec.~\ref{sec:toy} is only intended for the BR rather than for the amplitude. This is, however, enough for our purposes.}.
\item The single error given for the amplitude results in the second column of Table~\ref{tab:mainamp} corresponds to the error of the parameter $b_P$ in Eq.~\eqref{eq:BR} used for the TFF reconstruction. This parameter already includes a systematic error on top of the statistical one, which was estimated in Refs.~\cite{Escribano:2015nra,Escribano:2015yup} for $\eta$ and $\eta^{\prime}$, respectively.
\item The results collected in Table~\ref{tab:mainres} contain three different sources of error. The first corresponds to the often neglected uncertainty from BR$(P\rightarrow\gamma\gamma)$~\cite{Agashe:2014kda} in Eq.~\eqref{eq:BR}. This error is relevant for the $\eta$ and very important for the $\eta^{\prime}$. The second source is the one coming from $b_P$ as discussed in the previous item. The third is the systematic error from our method discussed in the previous section and given in Table~\ref{tab:c01res}, fourth column.
\end{itemize}
For completeness, Tables~\ref{tab:mainamp} and \ref{tab:mainres} also include the results for the $\pi^0\rightarrow e^+ e^- $ process~\cite{Masjuan:2015lca}.

\begin{table}
\centering
\footnotesize
\begin{tabular}{cclcl} \hline
Process & $\mathcal{A}(m_P^2)$ & $\mathcal{A}^{Z}(m_P^2)$ & \quad \quad $\mathcal{A}^{\textrm{app}}(m_P^2) $  \\ \hline
$\eta\rightarrow e^+e^-$       &  $(30.95\div31.51)(11) -21.92(0)i$        & $-0.03$  & $(27.53\div28.00)-21.92i$   \\ 
$\eta\rightarrow \mu^+\mu^-$   &  $-(1.52\div0.99)(5)  -5.47(0)i$         & $-0.03$ & $-(2.33\div1.87)-5.47i$   \\ \hline
$\eta^{\prime}\rightarrow e^+e^-$      &  $(47.4\div48.2)(5)  -(21.0)(5)i$     & $0.03$ & $(35.20\div35.66)-23.68i$  \\ 
$\eta^{\prime}\rightarrow \mu^+\mu^-$  &  $(2.95\div3.65)(19) -(6.27)(17)i$       & $0.03$  & $-(0.66\div0.20)-7.06i$  \\ \hline
$\pi^0\rightarrow e^+e^-$      &  $(10.00\div10.46)(12) -17.52i$           & $-0.05$ &   $(9.84\div10.30)-17.52i$   \\ \hline
\end{tabular}
\caption{Our results for the range $a_{P;11}\in(2b_P^2\div b_P^2)$, corresponding to (OPE$\div$Factorization). The error refers to the statistical error alone. We quote the $Z$-boson contribution $\mathcal{A}^{Z}(m_P^2)$ separately and quote the approximated $\mathcal{A}^{\textrm{app}}(m_P^2)$ calculation after expanding Eq.~(\ref{eq:loopint}) in terms of $m_l/m_P$ as well as $m_l/\Lambda$ and $m_P/\Lambda$. See details in the main text.}
\label{tab:mainamp}
\end{table}
\begin{table}
\footnotesize
\centering
\begin{tabular}{ccccc} \hline
Process & BR  & BR w/Z & BR app \\ \hline
$\eta\rightarrow e^+e^-$       & $(5.31 \div 5.44)(3)(2)(1)\times10^{-9}$ & $(5.32 \div 5.45)\times10^{-9}$ & $(4.58 \div 4.68)\times10^{-9}$  \\ 
$\eta\rightarrow \mu^+\mu^-$   &  $(4.72 \div 4.52)(2)(3)(4)\times10^{-6}$  & $(4.70 \div 4.51)\times10^{-6}$& $(5.16 \div4.88)\times10^{-6}$  \\ \hline
$\eta^{\prime}\rightarrow e^+e^-$      & $(1.82 \div 1.87)(7)(2)(16)\times10^{-10}$ & $(1.82 \div 1.87)\times10^{-10}$ & $(1.22 \div 1.24)\times10^{-10}$ \\ 
$\eta^{\prime}\rightarrow \mu^+\mu^-$  & $(1.36 \div 1.49)(5)(3)(25)\times10^{-7}$ & $(1.35 \div 1.48)\times10^{-7}$ & $(1.42 \div 1.41)\times10^{-7}$  \\ \hline
$\pi^0\rightarrow e^+e^-$      &  $(6.20\div6.35)(0)(4)(1)\times10^{-8}$  &$(6.22\div6.36)\times10^{-8}$ & $(6.17\div6.31)\times10^{-8}$  \\ \hline
\end{tabular}
\caption{Our results for the Branching Ratios for the range $a_{P;11}\in(2b_P^2\div b_P^2)$, corresponding to (OPE$\div$Factorization).  The errors refers to the statistical error for BR$(P\rightarrow\gamma\gamma)$, the error from $b_P$ and the systematic, respectively. We compare to the results either neglecting the $Z$-boson contribution (BR w/Z) or after expanding Eq.~(\ref{eq:loopint}) in terms of $m_l/m_P$ as well as $m_l/\Lambda$ and $m_P/\Lambda$ (BR app) discussed in the main text.}
\label{tab:mainres}
\end{table}
From those values it is clear that, for the $\eta$ decays, the most pressing task is the improvement of the double-virtual description which currently limits the theoretical uncertainty (the difference in the BRs between OPE and factorization is of about $0.1 \times 10^{-9}$ for the $ee$ channel, and about  $0.2 \times 10^{-6}$ for the $\mu \mu$ channel, far beyond the impact of the errors quoted). Notice that the systematic error of our method is of the order of the other error sources. To reduce our ignorance on the double virtual TFF, experimental data (or, eventually, lattice simulations) on the double-virtual TFF would be required.

Concerning the $\eta^{\prime}$, the situation is more complicated. In Table~\ref{tab:mainres}, third row, the largest errors arise from our systematic-error estimation in Table~\ref{tab:c01res}, while the errors coming from the normalization or the slope of the TFF are milder. The systematic uncertainty could be dramatically reduced if instead of estimating it using the toy model in Sec.~\ref{sec:toy}, we would look at the difference between the $C^0_1(Q_1^2,Q_2^2)$ and the $C^1_2(Q_1^2,Q_2^2)$ elements in the factorization approach (where no knowledge of the double virtuality is required). 
This is not surprising since, as emphasized previously, our toy model is not realistic enough and fails to describe the space-like region, which may hint an overestimation of the systematic uncertainty. 

In this respect, a more elaborated investigation including not only the low- and high-energy behaviors, but the information about the time-like region, such as physical resonances and threshold discontinuities, would be of interest in order to reach a similar precision to what is achieved in the $\eta$ case. Investigations in this respect are undergoing.

Our results may be compared to the experimental values given in Table~\ref{tab:exp}.
We find an interesting deviation in the $\eta\rightarrow\mu^+\mu^-$ channel. Still, the experimental accuracy prevents us from drawing any conclusion and a new experiment would be very welcomed. It has been recently suggested that this could be possible at the LHCb Collaboration~\cite{Huong:2016gob}. The result becomes even more interesting when comparing to the analogous $\pi^0\rightarrow e^+e^-$ anomaly (cf. our results in Table~\ref{tab:mainres} 
and the experimental value in Table~\ref{tab:exp}), as we find that, whereas for the $\pi^0$ case a very damped TFF at large energies was required to 
reproduce the experimental value~\cite{Masjuan:2015lca}, the $\eta$ case demands a smoothly falling TFF instead, which points to a puzzling situation.
Very interesting as well is the current bound on $\eta^{\prime}\rightarrow e^+e^-$, which is getting closer to the theoretical expectations. In this respect, it would be very stimulating to push for a new measurement. Such an effort is currently ongoing at VEPP-2000 $e^+e^-$collider at Novosibirsk, where they plan to increase their 
statistics by a factor of ten. Additionally, it has been suggested the possibility to measure $\eta'\to\mu^+\mu^-$ at LHCb~\cite{Huong:2016gob}.

The results above represent an important improvement with respect to previous studies. See for instance the comparison in Tables~\ref{tab:mainamp} and \ref{tab:mainres} between our exact results (second column) and the results obtained (last column) if we would have used the standard approximated calculation, widely used in the literature~\cite{Babu:1982yz,Bergstrom:1983ay,Dorokhov:2007bd}, which amounts to expand the kernel in Eq.~(\ref{eq:loopint}) in terms of $m_l/m_P$ as well as $m_l/\Lambda$ and $m_P/\Lambda$, with $\Lambda$ a cut-off of the loop integral (the hadronic scale driving the TFF), see Eq.~(3) in Ref.~\cite{Masjuan:2015lca}. Indeed, the missing corrections concerning the lepton mass along the lines of Ref.~\cite{Dorokhov:2009jd} are relevant for the $\mu^+\mu^-$ channel, which regretfully were not included in the recent Ref.~\cite{Dorokhov:2009xs}. 

Equally important is the correct implementation of both low- and high-energy TFF's behavior (i.e. using $b_P$ instead of the commonly employed resonance or VMD fit parameters as well as accounting for the OPE behavior), which for the $\eta$ case induces the largest effect. Beyond that,  the systematic error estimation is for the first time discussed and is by no means negligible.

On the contrary, the role of the $Z$ boson is almost negligible, though similar to some hadronic uncertainties. This is already a sign that new physics at the electroweak scale will not alter the results here presented and any eventual deviation from the Standard Model in these decays will demand a compelling New Physics benchmark.

\section{The low-energy description: the $\chi$PT approach}
\label{sec:chpt}

After reporting on our final results, we would like to discuss  the comments raised in the Introduction about the calculation of the pseudoscalar decays into lepton pairs within the framework of $\chi$PT. In this framework, the value for $\mathcal{A}(m_P^2)$ at leading order is given, in the $\overline{MS}$ renormalization scheme, as~\cite{Savage:1992ac,Knecht:1999gb}
\begin{equation}
\mathcal{A}^{\textrm{LO}}(m_{P}^2) = \frac{i\pi}{2\beta_{\ell}}L + 
\frac{1}{\beta_{\ell}} \left[ \frac{1}{4}L^2 
  +\frac{\pi^2}{12} + Li_2 \left(\frac{\beta_{\ell}-1}{1+\beta_{\ell}}\right)\right]
 -\frac{5}{2} +\frac{3}{2}\ln\left( \frac{m_{\ell}^2}{\mu^2} \right) + \chi(\mu),
\label{eq:loopaprox}
\end{equation}
where $L= \ln\left(\frac{1-\beta_{\ell}}{1+\beta_{\ell}}\right)$, $\beta_{\ell}$ is defined below Eq.~\eqref{eq:BR}, $\mu$ stands for the renormalization scale 
and $\chi(\mu)\equiv -(\chi(\mu)_1 + \chi(\mu)_2)/4$ is the (scale-dependent) counter-term. Given that, at leading order in $\chi$PT, $\chi(\mu)$ is common to all the $P\rightarrow\overline{\ell}\ell$ decays\footnote{This actually requires some assumptions when including the $\eta^{\prime}$, or assumptions to avoid the $\eta-\eta^{\prime}$ mixing if this is not included ---which is known to be extremely important to get the $P\rightarrow\gamma\gamma$ decays right. Invoking further assumptions, it has been shown that $K^0_L\rightarrow\overline{\ell}\ell$ decays 
could be used as well~\cite{GomezDumm:1998gw,Knecht:1999gb}.}, it has been argued~\cite{GomezDumm:1998gw} that any of the experimental values for these decays may be used to predict the rest.

To illustrate the accuracy of $\chi$PT at leading order, we want to quantify the role of the different approximations within $\chi$PT when calculating the parameter  $\chi(\mu)$ from Eq.~(\ref{eq:loopaprox}). For that purpose, we equate our final results for 
$\mathcal{A}(m_P^2)$ (see Table~\ref{tab:mainamp}) to the $\chi$PT result, Eq.~\eqref{eq:loopaprox}, from which $\chi(\mu)$ can be determined. The values obtained are displayed as $\chi(\mu)$ in Table~\ref{tab:chpt}  for $\mu=0.77$~GeV and represent the second main result in this work. In each entry on this table, we quote the extreme solutions given by the boundaries of the (OPE$\div$Fact) choices for $a_{P;1,1}$. The spread of values obtained for $\chi(\mu)$ in each channel manifest that the $\chi$PT predictions at this order are subject to significant corrections, which origin is detailed below.

First, there is a large splitting due to $U(3)$-breaking effects coming from the fact that $m_{\pi}<m_{\eta}<m_{\eta^{\prime}}$. To see the magnitude of them, we illustrate in 
Table~\ref{tab:chpt}, row $\chi(\mu)_{m_{\pi}}$, what would have been obtained for $\chi(\mu)$ if we would have imposed $m_{\eta^{\prime}}=m_{\eta}=m_{\pi}\equiv m_{\pi}$ in our $\mathcal{A}(m_P^2)$ calculations 
as well as in Eq.~\eqref{eq:loopaprox}. In this scenario, since $m_\pi< 2 m_\mu$, the decays into muon pairs cannot be calculated and are accounted for by a hyphen.

An additional difference arises from the different pseudoscalar meson transition form factors. Imposing that all of them are equal to the $\pi^0$ TFF, but the masses are restored to their real values, the results are collected under the row called $\chi(\mu)_{UV}$ in Table~\ref{tab:chpt}.

Finally, $m_{\ell}$ plays as well a relevant role in our decays. Corrections of the order ${\cal O}(m_{\ell}/m_P, \Lambda)$ are expected to arise dynamically, for instance, through pion loop effects as in Fig.~\ref{fig:UB} and implies that calculations at the next-to-leading order are relevant in $\chi$PT. This is specially important for the $\eta^{\prime}$, as this is the only mechanism able to generate an additional imaginary contribution within the $\chi$PT framework.
Moreover, this puts a word of caution on naive $\chi$PT analysis of these processes when looking for New Physics, as a different experimental extraction for $\chi(\mu)$ in either the electronic or muonic channel (for the same pseudoscalar) is expected within the Standard Model and does not necessarily imply lepton flavor violation. 
\begin{table}
\centering
\scriptsize
\begin{tabular}{cccccc} \hline
                                   & $\pi^0\rightarrow e^+e^-$ & $\eta\rightarrow e^+e^-$ & $\eta\rightarrow \mu^+\mu^-$ & $\eta^{\prime}\rightarrow e^+e^-$ & $\eta^{\prime}\rightarrow \mu^+\mu^-$ \\ \hline
    $\chi(\mu)$             & $(2.53\div2.99)$ & $(5.90\div6.46)$ & $(3.29\div3.82)$ & $(14.2\div14.9) + 2.52i$ & $(5.61\div6.31)+0.75i$ \\ 
$\chi(\mu)_{m_{\pi}}$ &  $(2.53\div2.99)$ & $(2.66\div3.12)$ & $-$ & $(2.16\div2.62)$ & $-$ \\  
    $\chi(\mu)_{UV}$ &  $(2.53\div2.99)$ & $(5.50\div6.05)$ & $(3.11\div3.64)$ & $(16.8\div17.7) + 7.09i$ & $(6.56\div7.35)+2.12i$ \\  \hline
\end{tabular}
\caption{Comparison between $\chi$PT counter-term $\chi(\mu)$ defined in Eq.~\eqref{eq:loopaprox} with its equal-mass version, $\chi(\mu)_{m_{\pi}}$, and the $U(3)$-symmetric version $\chi(\mu)_{UV}$. Results are always reported for the range (OPE$\div$Fact). See description in the main text.}
\label{tab:chpt}
\end{table}

\subsection{The $\pi^0$-exchange contribution to the $2S$ hyperfine-splitting in the muonic hydrogen}
\label{sec:Lshift}

The results collected in Table~\ref{tab:chpt}, first row, are also relevant for calculating the $\pi^0$-exchange contribution to the $2S$ hyperfine-splitting in the muonic hydrogen~\cite{Huong:2015naj}. Such calculation can be performed within $\chi$PT~\cite{Peset:2014jxa}, which involves again Eq.~\eqref{eq:loopaprox}. 
However, the kinematics of the process involve a vanishingly small $Q^2$ space-like momentum for the $\pi^0$ since the $\pi^0$-exchange contribution to the $2S$ hyperfine-splitting is a \textit{t-channel} exchange. As such, it is $\mathcal{A}(Q^2\simeq0)$ instead of $\mathcal{A}(m_{\pi}^2)$ which is relevant now~\cite{Huong:2015naj}, shifting the values obtained in Table~\ref{tab:chpt}. To illustrate this, we recalculate $\mathcal{A}(Q^2)$ from Eq.~\eqref{eq:loopint} taking the limit $Q^2 \to 0$, and obtain the new subtraction constant which should be used in Eq.~\eqref{eq:loopaprox} to reproduce our results. Using the $\pi^0$ TFF~\cite{Masjuan:2012wy} we obtain
\begin{equation}
\chi^{ee}_{\pi^0}(\mu)=(2.37\div2.83)
\label{eq:HFSe}
\end{equation}
for $\ell=e$, which is smaller than its counterpart collected in Table~\ref{tab:chpt}. However, for the $2S$ hyperfine-splitting in muonic hydrogen what is needed is the coupling to muons ($\ell=\mu$). In that case, we obtain\footnote{Our results Eqs.~\eqref{eq:HFSe} and \eqref{eq:HFSmu} do not change to the quoted digits when extrapolating to $\mathcal{A}(Q^2)$ 
up to energies $Q^2\sim(50)^2~\textrm{MeV}^2$, roughly at distances below $1.5\%$ the muonic Bohr radius.}
\begin{equation}
\chi^{\mu\mu}_{\pi^0}(\mu)=(2.18\div2.63),
\label{eq:HFSmu}
\end{equation}
which is even lower than Eq.~\eqref{eq:HFSe}. Note that the shift is of the order of the uncertainties quoted in Table~\ref{tab:mainamp} and arises again from the full $q^2$ and $m_{\ell}^2$ dependence in Eq.~\eqref{eq:loopint}, which is not accounted for at LO in $\chi$PT. We note that the relation between the $\chi^{ee}_{\pi^0}(\mu)$ in Eq.~\eqref{eq:HFSe} and $\chi^{\mu\mu}_{\pi^0}(\mu)$ in \eqref{eq:HFSmu} and the one extracted from the experimental results is non-trivial as it is TFF dependent. In quoting our results, we implicitly assume that there is no New-Physics contribution. However, if the current discrepancies among theory and experiment persists, indicating New Physics contribution ---which we will discuss in Sec.~\ref{sec:np}--- the connection between the experimental $\chi(\mu)$ and that in Eqs.~\eqref{eq:HFSe} and \eqref{eq:HFSmu} will depend on the particular scenario and will have to be reanalyzed.

The results above are illustrative as well regarding $(g-2)_{\mu}$ hadronic contributions, which in $\chi$PT involve $\chi(\mu)$ together with an additional counterterm, $C(\mu)$, as an input~\cite{RamseyMusolf:2002cy}. If we were able to determine $C(\mu)$ somehow, from $(g-2)_{e}$ for example, and $\chi(\mu)$ would be taken from the experimental $\pi^0\rightarrow e^+e^-$  result, extrapolating up to the $\mu$ case may imply a non-negligible error as illustrated above; similar effects may arise for $C(\mu)$ as well~\cite{Masjuan:2014rea,Masjuan:2015lca}.

\subsection{Higher order corrections}
\label{sec:cptcorr}

As discussed above, the precision which is reached at the LO in $\chi$PT for processes involving a $P\overline{\ell}\ell$ vertex may not be enough ---a feature which 
manifests when comparing the same process for a different $\ell=e,\mu$ channel. This suggests to look at the next-to-leading order. 
In this respect, $\chi$PT would yield a power series expansion for the TFF\footnote{For simplicity, we have assumed a single scale for the TFF inspired in typical VMD 
models. Note that logarithmic terms coming from loops are of course present too. However, they are subleading as compared to the power expansion and may be 
Taylor expanded for the $\pi^0$ and $\eta$ cases.} 
\begin{equation}
\label{eq:cpttff}
\tilde{F}_{P\gamma^*\gamma^*}(q_1^2,q_2^2) =  \underbrace{\phantom{\frac{1}{\Lambda^2}} \! \! \! \! \! \! \! \! \! 1}_{\textrm{LO}}
                                             +   \underbrace{ \frac{1}{\Lambda^2}(q_1^2 + q_2^2) }_{\textrm{NLO}} + 
                                              \underbrace{ \frac{1}{\Lambda^4}(q_1^4 + q_2^4) +   \frac{1}{\Lambda^4}(q_1^2q_2^2)  }_{\textrm{NNLO}} + \mathcal{O}\left(\frac{q^6}{\Lambda^6}\right)  .
\end{equation}
Then, we could calculate the result of \eqref{eq:loopint} for the TFF in \eqref{eq:cpttff}, 
\begin{align}
\mathcal{A}(q^2,m_{\ell}^2) &= \frac{2i}{\pi^2q^2}  \int  d^4k \frac{\left(k^2q^2-(k\cdot q)^2\right)}{k^2(q-k)^2\left((p-k)^2-m_{\ell}^2\right)}  \left[ 1 + \frac{(...)}{\Lambda^2}+ \frac{(...)}{\Lambda^4} + ...  \right] \nonumber \\
 \label{eq:CptExp}                           &\equiv   \mathcal{A}^{\textrm{LO}}(q^2,m_{\ell}^2)   +    \mathcal{A}^{\textrm{NLO}}(q^2,m_{\ell}^2)  +  \mathcal{A}^{\textrm{NNLO}}(q^2,m_{\ell}^2) + ... \ ,
\end{align}
where $\mathcal{A}^{\textrm{LO}}(q^2)$ has been given in \eqref{eq:loopaprox} and 
\begin{align}
\mathcal{A}^{\textrm{NLO}}(q^2,m_{\ell}^2)  \! &=  \frac{1}{3\Lambda^2}(q^2-10m_{\ell}^2)\left( 1 -L_{\ell}  \right) +\frac{1}{9\Lambda^2}(4m_{\ell}^2 - q^2) , \\
\mathcal{A}^{\textrm{NNLO}}(q^2,m_{\ell}^2)  \! &= \!  \! \left[\frac{126m{\ell}^4 -  \! q^4 -  \! 8m_{\ell}^2q^2 }{12\Lambda^4}L_{\ell}  
 + \frac{26m_{\ell}^2q^2 +  \! 7q^4 -  \! 702m_{\ell}^4}{72\Lambda^4} \right]  \!    , 
\end{align}
where $L_{\ell} = \ln(m_{\ell}^2/\Lambda^2)$\footnote{We note that, previous to the renormalization procedure, the loop integral produces 
$L_{\ell}=\ln(m_{\ell}^2/\mu^2)$ terms. It is after including the 
(non-explicitly shown) counterterms of the theory that the (scale independent) $L_{\ell} = \ln(m_{\ell}^2/\Lambda^2)$ 
result would appear. We emphasize that it is not our aim to perform a detailed higher-order $\chi$PT evaluation, rather than to illustrate the effects which are expected to appear.}.

We notice that the LO leading logs $L_{\ell}$ correspond ---not surprisingly as they arise from a power-like expansion as well--- 
to the corrections found in~\cite{Dorokhov:2008cd,Dorokhov:2009xs} if $\Lambda$ is taken as the VMD scale. We adopt then a much more modest approach and retain the leading 
logs alone, which represents a good approximation. This would produce a straightforward generalization to higher orders as well as a tool to estimate the convergence of the 
chiral expansion. Of particular relevance is the difference $\mathcal{A}(q^2,m_{e}^2)-\mathcal{A}(q^2,m_{\mu}^2)$, where one expects a better convergence of 
\eqref{eq:CptExp} due to partial cancellations. Taking into account the smallness of the electron mass, we find that such a shift is given as
\begin{multline}
\mathcal{A}(q^2,m_{e}^2)-\mathcal{A}(q^2,m_{\mu}^2) = \mathcal{A}^{\textrm{LO}}(q^2,m_{e}^2)-\mathcal{A}^{\textrm{LO}}(q^2,m_{\mu}^2) \\
  + \frac{q^2}{3\Lambda^2}\left( 1+ \frac{q^2}{4\Lambda^2} \right)  \ln\left( \frac{m_{\mu}^2}{m_e^2} \right) 
   + \frac{10m_{\mu}^2}{3\Lambda^2}\ln\left(\frac{\Lambda^2}{m_{\mu}^2} \right). \label{eq:NNLOcorr}
\end{multline} 
Whereas our theoretical results for the leptonic and muonic channels in Tab.~\ref{tab:mainamp} could not be reproduced at LO with an unique counterterm, the 
observed differences in Tab.~\ref{tab:chpt} and Section \ref{sec:Lshift} can be easily accounted for to a good approximation taking into account the additional  terms in \eqref{eq:NNLOcorr} 
---an exception is the $\eta'$ case, for which the pion loops cannot be neglected in order to extract an imaginary part.
The expansion above, Eq.~\eqref{eq:NNLOcorr}, proves extremely useful to relate different leptonic channels, which is not only relevant in the cases discussed above, 
but  for $\chi$PT studies on lepton flavor violation in $K_L\to \overline{\ell}\ell$ decays~\cite{Crivellin:2016vjc}.

\section{New physics contributions}
\label{sec:np}

\begin{figure}
   \includegraphics[width=\textwidth]{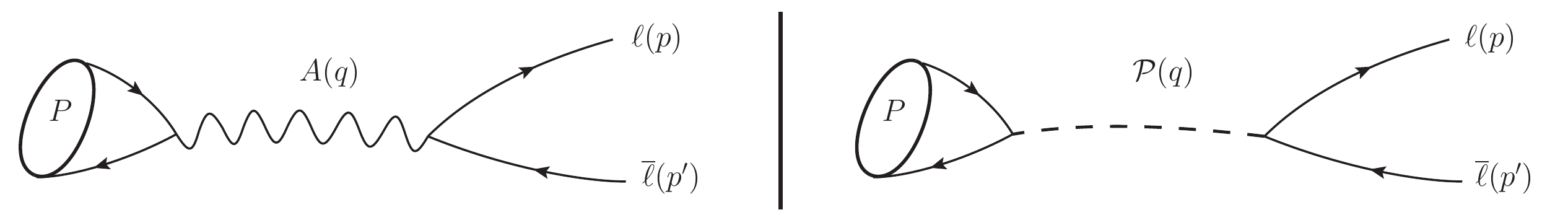}
   \caption{Left(right): additional tree level contributions from an axial(pseudoscalar) field. The $P$ within the blob 
                 stand for the pseudoscalar meson; $A(\mathcal{P})$ stands for the axial(pseudoscalar) field with momentum $q$; $\ell (\overline{\ell})$ for the (anti)lepton with momentum 
                 $p (p')$.\label{fig:np}}
\end{figure}

At this point, we are finally on a firm foot to discuss about possible New Physics (NP) 
contributions given the current discrepancies in the two existing measured decays. As discussed in~\cite{Masjuan:2015lca}, 
any additional contribution will always manifest, after Fierz-rearrangement, only through effective 
pseudoscalar $(\mathcal{P})$ and axial $(A)$ contributions, which given the existing well-motivated models, are conveniently expressed in the effective Lagrangian
\begin{equation}\nonumber
 \label{eq:NPL}
\mathcal{L} = \frac{g}{4m_W} \sum_{f} m_Ac^{A}_{f} \left(\overline{f}\slashed{A}\gamma_5f \right) + 2 m_f c^{\mathcal{P}}_{f} \left(\overline{f}i\gamma_5 f \right)\mathcal{P}, 
\end{equation}
where $g, m_W$ are the standard electroweak parameters, and $c^{A,\mathcal{P}}_{f}$ are dimensionless couplings to the fermions $f=\{u,d,s,e,\mu\}$.
These interactions yield additional tree-level contributions as shown in Fig.~\ref{fig:np}.
Their corresponding amplitudes read
\begin{align}
i\mathcal{M} = &\frac{igc^A_{\ell}m_A}{4m_W}[\overline{u}_{p,s}\gamma_{\mu}\gamma_5v_{p',s'}] \frac{-i\left(g_{\mu\nu} - \frac{q_{\mu}q_{\nu}}{m_A^2}\right)}{m_P^2-m_A^2}   \frac{igm_A}{4m_W} \overbrace{\sum_q \bra{0} c^A_q\overline{q}\gamma^{\mu}\gamma_5q\ket{P(q)}}^{\bra{0} J_{\mu5}^{\textrm{NP}}\ket{P(q)}}, \label{eq:npAX}\\
i\mathcal{M} = &\frac{i g c_{\ell}^{\mathcal{P}}}{2m_W} m_{\ell} [\overline{u}_{p,s}i\gamma_5 v_{p',s'}] \frac{i}{m_P^2 -M_{\mathcal{P}}^2} \frac{i g}{2m_W} 
\overbrace{ \sum_q \bra{0}c^{\mathcal{P}}_qm_q\overline{q}i\gamma_5q\ket{P(q)} }^{\bra{0} \mathcal{P}^{\textrm{NP}} \ket{P(q)} }, \label{eq:npPS}
\end{align}
for the axial and pseudoscalar contribution with masses $m_A$ and $M_{\mathcal{P}}$, respectively.

The hadronic matrix element $\bra{0} J_{\mu5}^{\textrm{NP}}\ket{P(q)}$ in Eq.~\eqref{eq:npAX} may be easily obtained using the octet-singlet basis or the flavor basis as an analogy to the SM axial current. From the pseudoscalar decay constants, defined in terms of the $U(3)$ axial currents\footnote{The singlet matrix is defined as $\lambda^1=\sqrt{2/3} \,   \mathds{1}$.},
\begin{equation}\nonumber
  \bra{0} J_{\mu5}^a \ket{P(q)} \equiv i q_{\mu}F_P^a, \qquad J_{\mu5}^a = \overline{q} \gamma_{\mu}\gamma_5 \frac{\lambda^a}{2} q, \quad q=(u,d,s)^T,
\end{equation}
where $a=8,1$ and $P=\eta, \eta^{\prime}$, we find that 
\begin{align}\nonumber
\bra{0} J^{\textrm{NP}}_{\mu5} \ket{P(p)}   = &  \sum_a \bra{0} \textrm{Tr}(J^{\textrm{NP}}_{\mu5}\lambda^a)J_{\mu5}^a \ket{P(p)} \nonumber \\ \nonumber
= & \sum_a \textrm{Tr}(\textrm{diag}(c^A_u,c^A_d,c^A_s)\lambda^a) \bra{0}J_{\mu5}^a\ket{P(p)}. \label{eq:axdec}
\end{align}
Inserting this equation back into Eq.~\eqref{eq:npAX} and using the spinor projector for the singlet state from Ref.~\cite{Martin:1970ai}, we obtain
\begin{align}\nonumber
i\mathcal{M} 
                     = & -ic_{\ell}^A\frac{G_F}{\sqrt{2}}m_{\ell}(\overline{u}_{p,s}i\gamma_5v_{p',s'})\sum_a\textrm{Tr}(J^{\textrm{NP}}_{\mu5}\lambda^a)F_P^a \\\nonumber
                     = & \, c^A_{\ell}m_{\ell}m_PG_F\sum_a\textrm{Tr}(J^{\textrm{NP}}_{\mu5}\lambda^a)F_P^a \, ,
\end{align}
where $G_F =\frac{\sqrt{2}g^2}{8m_W^2}$. This produces an additional contribution to Eq.~\eqref{eq:loopint} which reads
\begin{equation}\nonumber
\mathcal{A}(q^2) \rightarrow \mathcal{A}(q^2) + \frac{\sqrt{2}G_F}{4\alpha^2F_{P\gamma\gamma}} c^A_{\ell}\sum_a\textrm{Tr}(J^{\textrm{NP}}_{\mu5}\lambda^a)F_P^a.
\end{equation}
As an example, the $Z$-boson contribution is obtained after taking $c^Z_{u}=-c^Z_{d,s,e,\mu}=1$, leading for $P=\{\pi^0,\eta,\eta'\}$ 
\begin{equation}\nonumber
\mathcal{A}(q^2) \rightarrow \mathcal{A}(q^2) - \frac{2\sqrt{2}G_FF_{\pi}}{4\alpha^2F_{P\gamma\gamma}}\left\{1, \frac{F_{\eta}^8}{\sqrt{3}F_{\pi}} - \frac{F_{\eta}^1}{\sqrt{6}F_{\pi}} ,  \frac{F_{\eta^{\prime}}^8}{\sqrt{3}F_{\pi}} - \frac{F_{\eta^{\prime}}^1}{\sqrt{6}F_{\pi}}\right\} .
\end{equation}
Alternatively, we could have used the flavor basis instead~\cite{Escribano:2015yup}\footnote{Which amounts to trade $\lambda^8$ and $\lambda^1$ for $\lambda^q=\textrm{diag}(1,1,0)$ and $\lambda^s=\textrm{diag}(0,0,\sqrt{2})$.}, then, 
\begin{equation}\nonumber
\mathcal{A}(q^2) \rightarrow \mathcal{A}(q^2) - \frac{2\sqrt{2}G_FF_{\pi}}{4\alpha^2F_{P\gamma\gamma}}\left\{1, -\frac{F_{\eta}^s}{\sqrt{2}F_{\pi}} , -\frac{F_{\eta^{\prime}}^s}{\sqrt{2}F_{\pi}}\right\} .
\end{equation}

For the pseudoscalar contribution, the $\bra{0} \mathcal{P}^{\textrm{NP}} \ket{P(q)} $ hadronic matrix element determination in Eq.~(\ref{eq:npPS}) is more complicated whenever the singlet component is involved. This is the case for both $\eta$ and $\eta^{\prime}$ as they are an admixture of the octet and singlet states. To illustrate this, let us calculate such matrix elements at leading order (LO) in $\chi$PT, which amounts to retain the leading pseudoscalar $\mathcal{P}\times P$ term arising from the interaction between the pseudoscalar field $P$ and the pseudoscalar 
current  $\mathcal{P}$ defined in $\chi$PT from the building block $\chi \equiv 2 B_0i \mathcal{P}$ ($B_0$ is the low-energy constant related to the scalar singlet quark condensate $\langle \bar{q}q\rangle_0$ in the chiral limit). Then, in the presence of NP of pseudoscalar type, $\chi \rightarrow2B_0i \mathcal{P}^{\textrm{NP}}$. For the $\pi^0$ such term corresponds to 
\begin{equation}\nonumber
 F_0B_0\hat{m}(c_u^{\mathcal{P}} - c_d^{\mathcal{P}})\mathcal{P}^{\textrm{NP}}\pi^0, 
\end{equation}
with $\hat{m} = (m_u +m_d)/2$, from which the matrix element reads ($2B_0 \hat{m} =  m_\pi^2$)
\begin{equation} \nonumber
\label{eq:pspion}
\bra{0} \mathcal{P}^{\textrm{NP}}\ket{\pi^0}=  F_0B_0\hat{m}(c_u^{\mathcal{P}} - c_d^{\mathcal{P}}) = \frac{F_{\pi}}{2}m_{\pi}^2(c_u^{\mathcal{P}} - c_d^{\mathcal{P}}).
\end{equation}
For the $\eta$ and $\eta^{\prime}$ such term is more involved and at LO reads
\begin{equation}\nonumber
\label{eq:psmxel1}
F_0B_0\mathcal{P}^{\textrm{NP}}\left(\frac{1}{\sqrt{3}}\left( \hat{m}(c^{\mathcal{P}}_u+c^{\mathcal{P}}_d) - 2c^{\mathcal{P}}_s m_s \right)\eta_8 + \sqrt{\frac{2}{3}}\left( \hat{m}(c^{\mathcal{P}}_u+c^{\mathcal{P}}_d) + c^{\mathcal{P}}_s m_s \right)\eta_1 \right).
\end{equation}
After relabeling, introducing $g_8\equiv (c^{\mathcal{P}}_u+c^{\mathcal{P}}_d-2c^{\mathcal{P}}_s)/\sqrt{3}$ and $g_1\equiv\sqrt{2}(c^{\mathcal{P}}_u+c^{\mathcal{P}}_d+c^{\mathcal{P}}_s)/\sqrt{3}$, it can be expressed as
\begin{equation}\nonumber
\frac{F_0}{2}\left( \eta_8(g_8M_8^2+g_1M_{81}^2) + \eta_1(g_8M_{81}^2+g_1M_1^2) \right)\mathcal{P}^{\textrm{NP}},
\end{equation}
where $M_8^2,M_1^2$ and $M_{81}^2$ are defined in Refs.~\cite{Escribano:2010wt,Bickert:2015cia}.
Finally, using the $\eta-\eta^{\prime}$ masses, mixing and decay constants at LO\footnote{At LO, $\eta_{8(1)}=\eta(\eta^{\prime})\cos\theta_P \pm \eta^{\prime}(\eta)\sin\theta_P$, $\theta_P=-19.6^{\circ}$~\cite{Bickert:2015cia} 
and the decay constants read $F_{\eta}^8 = F_0 \cos\theta_P, F_{\eta^{\prime}}^8 = F_0 \sin\theta_P, F_{\eta}^1 = F_0\sin\theta_P, F_{\eta^{\prime}}^1 = F_0 \cos\theta_P$. In addition, $\sin (2\theta_P) = \frac{2M_{81}^2}{M_{\eta^{\prime}}^2-M_{\eta}^2}$, see Ref.~\cite{Bickert:2015cia}.}, 
we obtain for the matrix element
\begin{equation}
\label{eq:cptps}
\bra{0} \mathcal{P}^{\textrm{NP}} \ket{\eta(\eta^{\prime})}= \sum_a \frac{1}{2}F_{\eta(\eta^{\prime})}^a g_a m_{\eta(\eta^{\prime})}^2\left(1-\delta^{a1}\frac{M_0^2}{M_{\eta(\eta^{\prime})}^2} \right),
\end{equation}
where $g_a$ has been defined above and $M_0^2=6\tau/F_0^2$ is the topological mass term~\cite{Bickert:2015cia}. After some algebra, we have obtained 
a relation which is very similar 
to the $\pi^0$ result ---except for the singlet $a=1$ term--- and resembling that of the axial current matrix element. 
The previous calculation applies so far up to LO. However, such precision is not enough to reproduce the observed $\eta-\eta'$ mixing, which requires higher order calculations. 
In order to provide a general result, valid at any order and related to known parameters, it is convenient to recall the Ward identity~\cite{Bickert:2015cia}
\begin{equation}\nonumber
\partial^{\mu}J_{\mu5}^{a} = \left\{ \mathcal{P}^a,\mathcal{M} \right\} + \delta^{a1}\sqrt{N_F/2}~\omega; \quad \mathcal{P}^a = \overline{q}i\gamma_5\frac{\lambda^a}{2}q, \ \  q=(u,d,s)^T,
\end{equation}
where $\mathcal{M}=\textrm{diag}(\hat{m},\hat{m},m_s)$ is the quark mass matrix. In such a way, the pseudoscalar current may be expressed in terms of the axial 
current and the winding number density $\omega$. Then, using the same algebra as previously, the matrix element can be expressed as  
\begin{align}\nonumber
\bra{0} \mathcal{P}^{\textrm{NP}}\ket{P(p)} = &\frac{1}{2} \sum_a\textrm{Tr}(\mathcal{P}^{\textrm{NP}}\lambda^a) \bra{0}\partial^{\mu}J_{5\mu}^a -\delta^{a1}\sqrt{3/2}~\omega \ket{P(p)} \\
 = & \frac{m_P^2}{2} \sum_a\textrm{Tr}(\textrm{diag}(c_u^{\mathcal{P}},c_d^{\mathcal{P}},c_s^{\mathcal{P}})\lambda^a)F_{P}^a(1-\Delta\delta^{1a}), \label{eq:singapp}
\end{align}
where $\Delta=\bra{0}\sqrt{3/2}~\omega \ket{P}/m_P^2F_P^1$. Still, $\Delta$ needs to be determined. Neglecting the $u$ and $d$ quark masses, $\omega$ may be expressed as~\cite{Escribano:2005qq}:
\begin{equation}\nonumber
\sqrt{3/2}~\omega = \partial^{\mu}J_{\mu5}^1 + \frac{1}{\sqrt{2}}\partial^{\mu}J_{\mu5}^8 .
\end{equation}
Plugging this relation into (\ref{eq:singapp}), we obtain $\Delta=1+F_P^8/(\sqrt{2}F_P^1)$, so the pseudoscalar contribution to $P\rightarrow \overline{\ell}\ell$ can be finally expressed as
\begin{align}\nonumber
i\mathcal{M} = & -i [\overline{u}_{p,s}i\gamma_5 v_{p',s'}]\frac{G_F}{\sqrt{2}} \frac{m_P^2m_{\ell}c^{\mathcal{P}}_{\ell}}{m_P^2-M_{\mathcal{P}}^2} 
\sum_a\textrm{Tr}(\mathcal{P}^{\textrm{NP}}\lambda^a)F_{P}^a\left(1-\delta^{1a}(1+\frac{F_P^8}{F_P^1\sqrt{2}})\right),\\
= & \, G_F \frac{m_P^3m_{\ell}c^{\mathcal{P}}_{\ell}}{m_P^2-M_{\mathcal{P}}^2} 
\sum_a\textrm{Tr}(\mathcal{P}^{\textrm{NP}}\lambda^a)F_{P}^a\left(1-\delta^{1a}(1+\frac{F_P^8}{F_P^1\sqrt{2}})\right).
\end{align}
This induces an additional contribution to the $\mathcal{A}(q^2)$ loop amplitude in Eq.~\eqref{eq:loopint},
\begin{equation}\nonumber
\mathcal{A}(q^2) \rightarrow \mathcal{A}(q^2) + \frac{\sqrt{2}G_Fm_P^2 c^{\textrm{NP}}_{\ell}}{4\alpha^2F_{P\gamma\gamma}(m_P^2-M^2_{\mathcal{P}})}\sum_a\textrm{Tr}(\mathcal{P}^{\textrm{NP}}\lambda^a)F_{P}^a\left(1-\delta^{1a}\left(1+\frac{F_P^8}{F_P^1\sqrt{2}}\right)\right).
\end{equation}
We note that the approximation taken for calculating the $\bra{0}\omega\ket{P}$ matrix element has been used with great success in 
$J/\Psi\rightarrow\gamma\eta(\eta^{\prime})$ decays~\cite{Escribano:2005qq}. Actually, at LO in $\chi$PT\footnote{At this order, the $\eta$ and $\eta^{\prime}$ masses 
are~\cite{Bickert:2015cia} $M_{\eta}^2=0.244\textrm{GeV}^2$ $M_{\eta^{\prime}}^2=0.917\textrm{GeV}^2$ and 
$M_{0}^2=0.673\textrm{GeV}^2$.}, the difference between Eq.~\eqref{eq:cptps} and Eq.~\eqref{eq:singapp} is of $8\%(1\%)$ for the $\eta(\eta^{\prime})$, precise enough for our study.
The nice feature from this approach is to provide an appropriate description for the singlet component valid at any order in the chiral expansion and based on known parameters ---as long as the approximation employed holds to the required precision. 

In the flavor basis, neglecting the $u$ and $d$ quark masses, only the strange part contributes. Using an analogous procedure, we find
\begin{equation}\nonumber
\mathcal{A}(q^2) \rightarrow \mathcal{A}(q^2) + \frac{\sqrt{2}G_Fm_P^2 c^{\textrm{NP}}_{\ell}}{4\alpha^2F_{P\gamma\gamma}(m_P^2-M^2_{\mathcal{P}})}\sum_a\textrm{Tr}(\mathcal{P}^{\textrm{NP}}\lambda^a)F_{P}^a\left(1-\frac{\delta^{as}F_P^q}{\sqrt{2}F_P^s}\right)(1-\delta^{aq}).
\end{equation}
All in all, both contributions may be summarized to yield an additional term modifying Eq.~(\ref{eq:loopint}) as
\begin{equation}
   \label{eq:NP}
   \mathcal{A}(q^2) \rightarrow  \mathcal{A}(q^2) + \frac{\sqrt{2} G_F F_{\pi}}{4\alpha_{em}^2 F_{P\gamma\gamma}}(\lambda^A_P+\lambda^{\mathcal{P}}_P) , 
\end{equation}
where $G_F$ is the Fermi coupling constant, and $F_{\pi}\simeq92$~MeV is the pion decay constant. The $\lambda$-terms depend on the pseudoscalar meson structure, 
which for the $\eta$ and $\eta^{\prime}$ involve the mixing parameters. In the flavor-mixing scheme, they read\footnote{By definition, $F_{\pi^0}^8=F_{\pi^0}^1\equiv0$ and 
$F_{\pi^0}^3\equiv F_{\pi}$. From Ref.~\cite{Escribano:2015yup}, $F_{\eta(\eta^{\prime})}^q=0.84(0.72)F_{\pi}$, $F_{\eta(\eta^{\prime})}^s=-0.90(1.14)F_{\pi}$ 
and $F_{\eta(\eta^{\prime})}^3\equiv0$. See details in~\cite{Escribano:2015yup} for the errors on these parameters.}
\begin{align}
\label{eq:NPA}
\lambda^{A}_{P} & =  c^A_{\ell} \left[    \frac{F_{P}^3}{F_{\pi}} \left( c^{A}_u - c^{A}_d \right)   +   \frac{F_{P}^q}{F_{\pi}} \left( c^{A}_u + c^{A}_d \right)    +    \frac{F_{P}^s}{F_{\pi}} \sqrt{2}c^{A}_s \right], \\ 
\label{eq:NPP}
\lambda^{\mathcal{P}}_{P} & =   \frac{c^{\mathcal{P}}_{\ell}}{1- \frac{M_{\mathcal{P}}^2}{m_{P}^2}} 
 \left[    \frac{F_{P}^3}{F_{\pi}} \left( c^{\mathcal{P}}_u - c^{\mathcal{P}}_d \right)   +   \frac{F_{P}^q}{F_{\pi}} \left( -c^{\mathcal{P}}_s \right)    +    \frac{F_{P}^s}{F_{\pi}} \sqrt{2}c^{\mathcal{P}}_s \right].  
\end{align}
Taking the result from the mixing parameters in Ref.~\cite{Escribano:2015nra} to numerically calculate Eqs.~(\ref{eq:NPA}, \ref{eq:NPP}), Eq.~\eqref{eq:NP} yields
\begin{align}\nonumber
   \label{eq:NPnumeric}
   & \mathcal{A}(m_{\pi^0}^2) + 0.026 \left(c^A_{\ell}(c^A_u -c^A_d) + c^{\mathcal{P}}_{\ell}(c^{\mathcal{P}}_u -c^{\mathcal{P}}_d)(1-M_{\mathcal{P}}^2/m_P^2)^{-1}  \right) , \\\nonumber
   & \mathcal{A}(m_{\eta}^2) + 0.026  \left(0.84c^A_{\ell}(c^A_u +c^A_d) - 1.27c^A_{\ell}c^A_s  -2.11c^{\mathcal{P}}_{\ell}c^{\mathcal{P}}_{s}(1-M_{\mathcal{P}}^2/m_P^2)^{-1} \right) , \\\nonumber
   & \mathcal{A}(m_{\eta^{\prime}}^2) + 0.021 \left(0.72c^A_{\ell}(c^A_u +c^A_d) + 1.61c^A_{\ell}c^A_s  +0.89c^{\mathcal{P}}_{\ell}c^{\mathcal{P}}_{s}(1-M_{\mathcal{P}}^2/m_P^2)^{-1} \right) . 
\end{align}
To discuss the sensitivity of each particular channel to NP, it is convenient to cast a very approximate result for $\mathcal{A}(m_P^2)$, namely 
\begin{equation}
\label{eq:AmpApp}
\mathcal{A}(m_P^2) \simeq i\pi \left[ \ln\left(\frac{m_{\ell}}{m_P}\right) \right] + \left[ \ln^2\left(\frac{m_{\ell}}{m_P}\right)  - 3\ln\left(\frac{\Lambda}{m_{\ell}}\right) +\delta_{\textrm{NP}} \right],
\end{equation}
where $\Lambda$ is an effective hadronic scale characterizing the TFF and $\delta_{\textrm{NP}}$ is the NP contribution in Eq.~\eqref{eq:NP}. 
From Eq.~\eqref{eq:AmpApp}, we see that, as the lepton mass gets lighter, the amplitude will be dominated by the $\ln(m_{\ell}/m_P)$ terms, which become large and 
make the NP contribution harder to see. Indeed, for $\ell = e$, the relative NP contribution to the BR is approximately given by 
$ 2 \delta_{\textrm{NP}}(\ln^2(\frac{m_{e}}{m_P}) + \pi^2)^{-1}$. If we are aiming to find contributions from NP, it is therefore much easier to look for the $\ell=\mu$ 
channel as the NP part is insensitive to $m_{\ell}$ (see Eq.~\eqref{eq:NP}). 

With respect to $m_P$, from the logarithmic scaling, we infer that there is no big difference in the SM in choosing either $\pi^0,\eta$, or $\eta^{\prime}$ as their masses are of same order. Furthermore, the NP axial contribution does not see the $m_P$, meaning that is equally likely to appear in any case. This contrasts with the pseudoscalar NP contribution, which strongly depends on $m_P$ (cf. Eq.~\eqref{eq:NPP}) and gets bigger as $m_P$ and $M_{\mathcal{P}}$ (the mass of the new pseudoscalar particle) approach each other. Still, this is \textit{a priori} irrelevant unless there is a well-motivated NP scale which is close to either the $\pi^0, \eta$, or $\eta^{\prime}$ masses.

From this discussion, we conclude that $\eta(\eta^{\prime})\rightarrow \mu^+\mu^-$ decays are the best candidates to look for NP effects (as the $\pi^0$ cannot decay into muons). For illustrating the statements above, we give the approximate NP contribution to the branching ratio for each particular process,
\begin{align}\nonumber
   \label{eq:NPnumeric}
   & BR(\pi^0\rightarrow e^+e^-)\left( 1 +0.001\left[c^A_{\ell}(c^A_u -c^A_d) +  c^{\mathcal{P}}_{\ell}\frac{c^{\mathcal{P}}_u -c^{\mathcal{P}}_d}{1-M_{\mathcal{P}}^2/m_P^2} \right]\right) , \\\nonumber
   & BR(\eta\rightarrow^{\mu^+\mu^-}_{e^+e^-}) \left(1 +\left(^{-0.002}_{+0.001}\right) \left[ 0.84c^A_{\ell}(c^A_u +c^A_d) - 1.27c^A_{\ell}c^A_s  -\frac{2.11c^{\mathcal{P}}_{\ell}c^{\mathcal{P}}_{s}}{1-M_{\mathcal{P}}^2/m_P^2} \right]\right) , \\\nonumber
   & BR(\eta^{\prime}\rightarrow^{\mu^+\mu^-}_{e^+e^-}) \left(1 +\left(^{+0.003}_{+0.001}\right)\left[ 0.72c^A_{\ell}(c^A_u +c^A_d) + 1.61c^A_{\ell}c^A_s  + \frac{0.89c^{\mathcal{P}}_{\ell}c^{\mathcal{P}}_{s}}{1-M_{\mathcal{P}}^2/m_P^2} \right]\right) . 
\end{align}
We see that, as stated above, the $\ell = e$ channel has the same sensitivity for every pseudoscalar. For $\ell=\mu$ we find that for $\eta(\eta^{\prime})$ is two(three) times more sensitive as the $\ell=e$ channel. These numbers imply, together with the experimental precision reached for the $\pi^0(\eta)$ decay (we do not consider the central value, but the obtained precision), bounds for the $c^{A}$ parameters of the order $c^A\sim7(8)$. As an example, for the $Z$ boson $(c^A_{\ell}=c^A_{d,s}=-c^A_u\equiv1)$, the $c^A_f$ combination is $-2(-1.27)(1.61)$ for 
$\pi^0(\eta)(\eta^{\prime})$.

Interesting enough, a typical $Z$-like contribution has opposite sign for $\pi^0\rightarrow e^+e^-$ than for $\eta\rightarrow\mu^+\mu^-$, contrary to experimental implications. This would suggest either different couplings (necessarily $SU(2)$ breaking), or lepton flavor violating (LFV) models, which would couple different to distinct generation of quarks, leptons, or both. Moreover, in order to avoid $(g-2)_{\mu}$ problems, we would need either some balance from an additional vector-like contribution\footnote{The dominant Schwinger-like contribution for a vector(scalar)-like coupling has positive sign whereas the axial(pseudoscalar) one has opposite sign, providing a fine tuning cancelation.} or, again, LFV models in which the coupling to the muon is suppressed.

For a pseudoscalar contribution, as in Ref.~\cite{Chang:2008np}, the effective couplings may become even larger as the new particle mass approaches the $\pi^0,\eta,\eta^{\prime}$ masses, meaning that would be visible for one of the pseudoscalars alone. Finally, we comment on the existing correlations given the pseudoscalar structure. We see for instance that $\pi^0\rightarrow e^+e^-$ and $\eta\rightarrow \mu^+\mu^-$ are, in general, anti-correlated unless there is a pseudoscalar particle $\mathcal{P}$ with $m_{\pi^0}<m_{\mathcal{P}}<m_{\eta}$ (or a different structure for distinct generations). Again, $(g-2)_{\mu}$ would play an important constraint for the pseudoscalar case as well.

To conclude, there is still the chance to look for NP contributions, specially in the $\ell=\mu$ channel, and a variety of phenomenology is possible depending on which kind of interaction is chosen. Still, our study suggests to go beyond simple scenarios. This seems nevertheless the standard in high energy physics nowadays, and scenarios of this kind have been and are still studied at present. In this discussion, we have omitted a detailed discussion of available physical constraints for these scenarios. This constitutes a field of study by itself~\cite{Davoudiasl:2012ag,Davoudiasl:2012qa,Chang:2008np,Carlson:2012pc,Karshenboim:2014tka}.

\section{Conclusions}
\label{ref:conclusions}

In this work, we discussed a novel approach for evaluating the Standard Model prediction for $\eta(\eta^{\prime})\rightarrow\overline{\ell}\ell$ decays, which are mainly driven by the non-perturbative regime of QCD. This was made possible using the machinery of Canterbury approximants (CA), an extension of Pad\'e approximants to the double virtual case. This approach is data driven, systematic and allows for the correct implementation of the low- and high energy QCD requirements, which are key points in the calculation. In addition, our method implements a systematic error and the results come from a full numerical evaluation for the loop integral, which were not included in most of the previous approaches. From our experience in~\cite{Masjuan:2015lca}, we expect that higher elements in the approximants sequence yield a value very similar to the OPE $a_{P,1,1}=2b_P^2$ choice which also encodes the appropriate high-energy behavior, which means only the correct $Q^{-2}$ dependence but not its exact coefficient. We quote these as our final number, incorporating the difference with respect to the factorization $a_{P,1,1}=b_P^2$ choice as an asymmetric systematic error in Table~\ref{tab:fin}. Given the current experimental errors, the achieved precision is accurate enough and does not require the computation of higher elements in the CA sequence, which poses a major complication with respect to the $\pi^0$ case. We postpone such a calculation for the near future in case new experiments prove to require a higher precision.
\begin{table}[h]
\centering
\begin{tabular}{cc} \hline
Process  & BR \\ \hline
$\eta \to e^+e^-$  & $5.31(^{+14}_{-4})\times10^{-9}$ \\
$\eta \to \mu^+\mu^-$  & $4.72(^{+5}_{-21})\times10^{-6}$ \\
$\eta' \to e^+e^-$  & $1.82(^{+18}_{-18})\times10^{-10}$ \\
$\eta' \to \mu^+\mu^-$  & $1.36(^{+29}_{-26})\times10^{-7}$ \\ \hline
\end{tabular}
\caption{Our final results for the BRs. We take as the central value the OPE result and add as an additional (asymmetric) systematic error the difference with respect to the factorization choice.\label{tab:fin}}
\end{table}

In this work, we have found a remarkable feature of our approach, namely, that by abandoning the old conceptions on rational approaches ---where the poles are associated to physical resonances and fixed in advance--- in favor of the Pad\'e Theory point of view ---where poles are effective parameters accounting for the analytic structure of the underlying function--- it is possible to effectively reproduce non-trivial effects encoded in the TFF, such as branch cuts, which is the reason for which we can perform a calculation for the $\eta(\eta^{\prime})\rightarrow\overline{\ell}\ell$ decays, beyond the possibilities of resonance approaches.

We have found that, similar to the $\pi^0\rightarrow e^+e^-$ case, there is an anomaly for the $\eta\rightarrow\mu^+\mu^-$ decay, which would make a new measurement very interesting.

In addition, we have discussed the difficulties in achieving a precise description for these processes from $\chi$PT, though it would be \textit{a priori} an adequate tool for this purpose; in this respect, we have parametrized the expected higher-order corrections for further studies, which could be useful for LFV studies in $K_L\to\bar{\ell}\ell$. As an outcome, we have parameterized the impact of the $\pi^0$-exchange contribution to the $2S$ hyperfine-splitting in the muonic hydrogen based on our data-driven results.

Finally, we have discussed the implications of New Physics scenarios, which should effectively be of either axial or pseudoscalar nature. Our results suggest that such contributions should arise either from $SU(2)$ isospin-breaking couplings or lepton flavor violating scenarios.

\appendix
\section{Generalities of Canterbury approximants}
\label{sec:app}

In this appendix, we define and illustrate the performance of Canterbury approximants specializing to the case of symmetric functions $f(x,y)=f(y,x)$. Given  
the original function formal expansion
\begin{equation}
f(x,y) = \sum_{\alpha,\beta}c_{\alpha,\beta}x^{\alpha}y^{\beta}, \quad (c_{\alpha,\beta} = c_{\beta,\alpha}),
\end{equation}
the Canterbury approximants $C^N_M(x,y)$ are rational functions of the bivariate degree $N$ and degree $M$ polynomials, $R_N(x,y)$ and $Q_M(x,y)$\footnote{Note that the polynomials 
are constructed as to have the maximum power in each variable rather than a total maximum power in $x^iy^j$ with $i+j\leq N(M)$~\cite{Chisholm}. In other words, 
$R_N(x,y)$ contains a term $x^Ny^N$ and $(N+1)(N+2)/2$ terms in total.}, respectively,
\begin{equation}
\label{eq:ca}
C^N_M(x,y) =  \frac{R_N(x,y)}{Q_M(x,y)} =  \frac{\sum_{k,l=0}^{N}  a_{k,l}x^ky^l}{\sum_{k,l=0}^{M}  b_{i,j}x^iy^j}, 
\end{equation}
where the coefficients ($i\geq j$) $a_{i,j}\in\mathcal{N}$ and $b_{i,j}\in\mathcal{D}$ are determined by the accuracy-through-order conditions ($b_{0,0}=1$ as a part of the definition)
\begin{align}
  \sum_{i=0}^{\alpha}\sum_{j=0}^{\beta} b_{i,j} c_{\alpha-i,\beta-j} &= a_{\alpha,\beta} \quad  \textrm{for } (\alpha,\beta)\in\mathcal{N},  \label{eq:cadeq1} \\
   \sum_{i=0}^{\textrm{min}(\alpha,M)}\sum_{j=0}^{\textrm{min}(\beta,M)}b_{i,j} c_{\alpha-i,\beta-j} &= 0  \quad  \textrm{for } (\alpha,\beta)\in\mathcal{E},  (\alpha,\beta)\notin\mathcal{N}, \label{eq:cadeq2}
\end{align}
where $\textrm{dim}(\mathcal{E}) = \textrm{dim}(\mathcal{N}) + \textrm{dim}(\mathcal{D}) -1$.

\subsection{Performance of Canterbury approximants with the help of toy models}

\begin{figure}
\centering
  \includegraphics[width=0.325\textwidth]{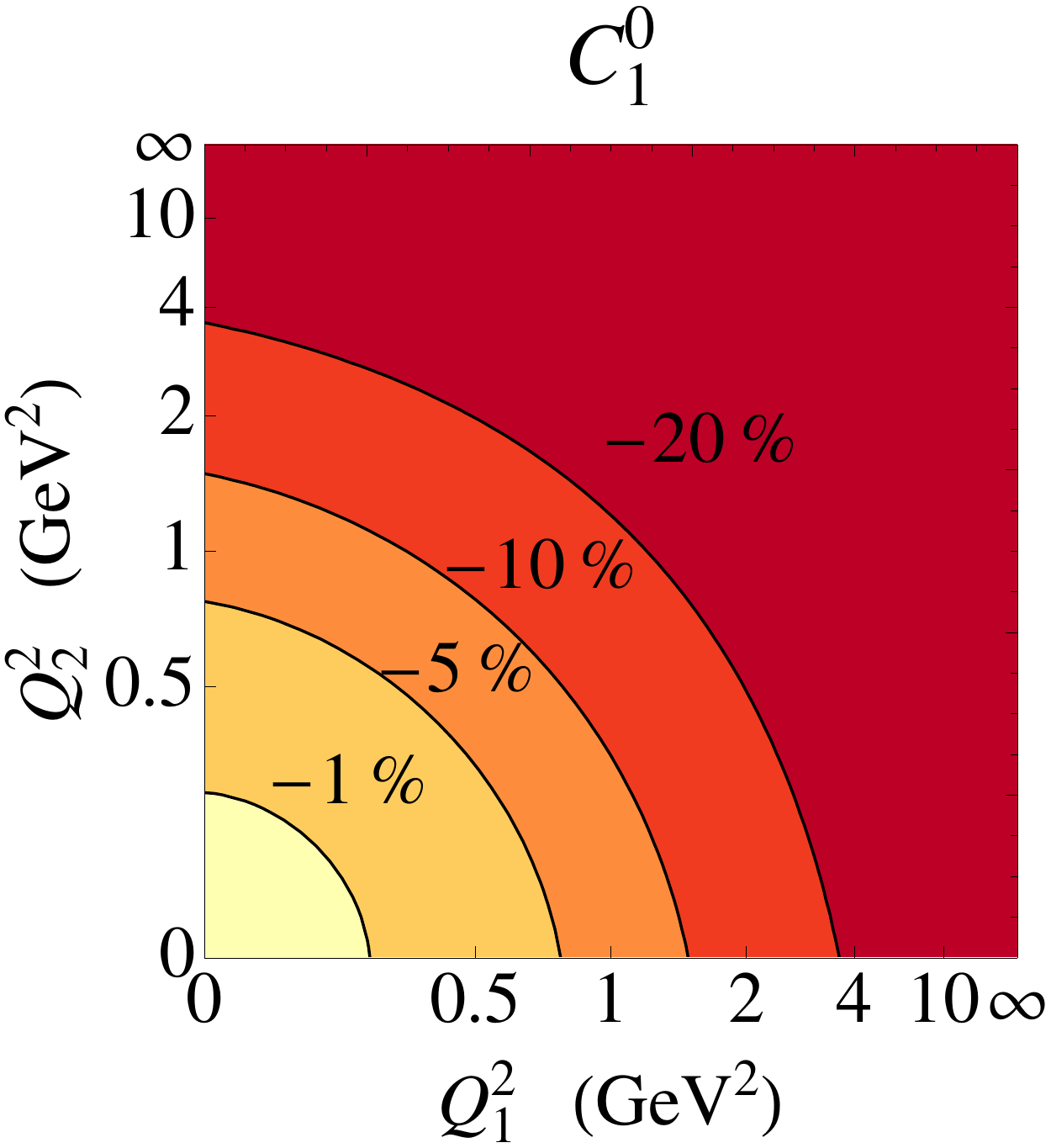}
  \includegraphics[width=0.325\textwidth]{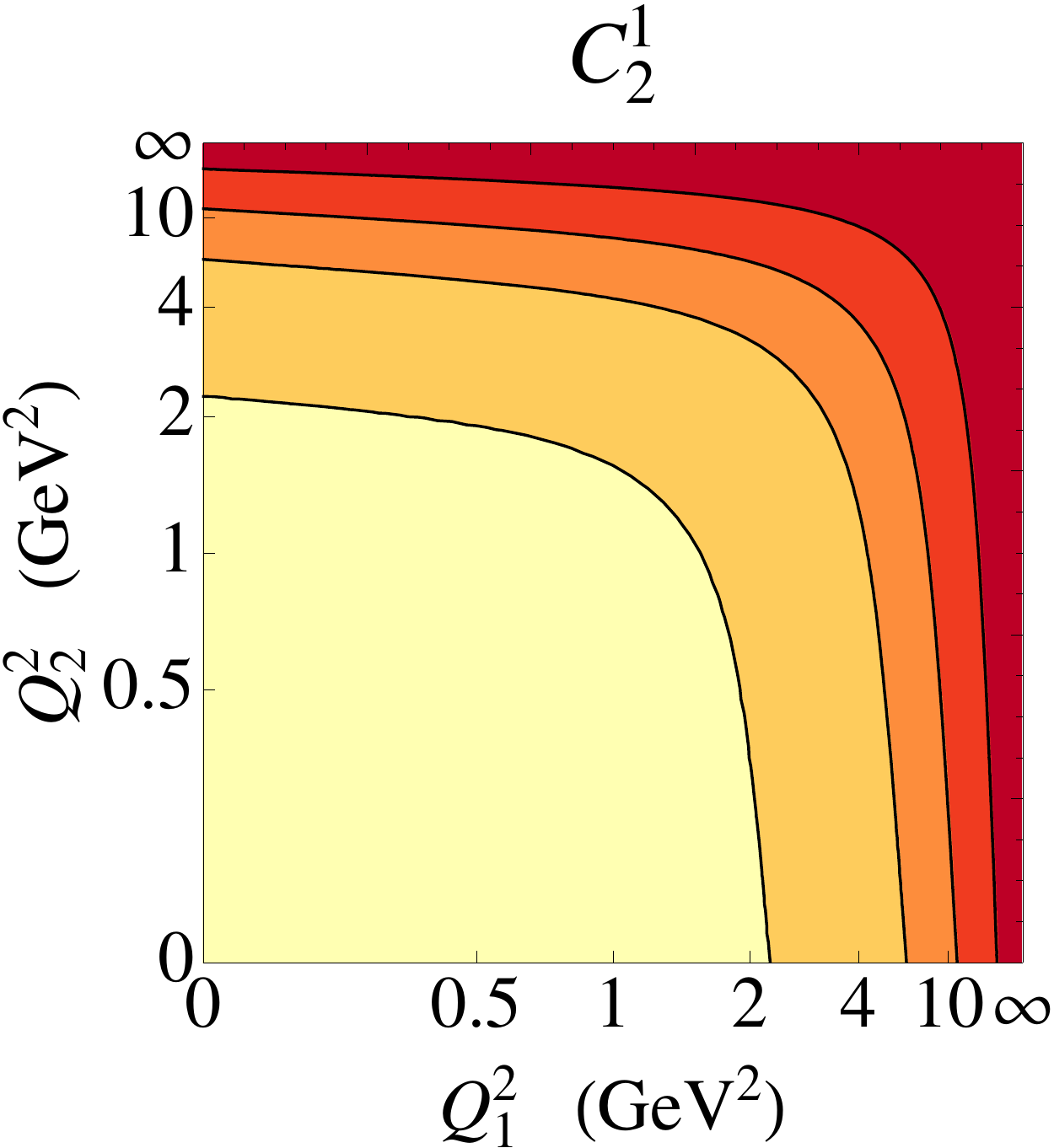}
  \includegraphics[width=0.325\textwidth]{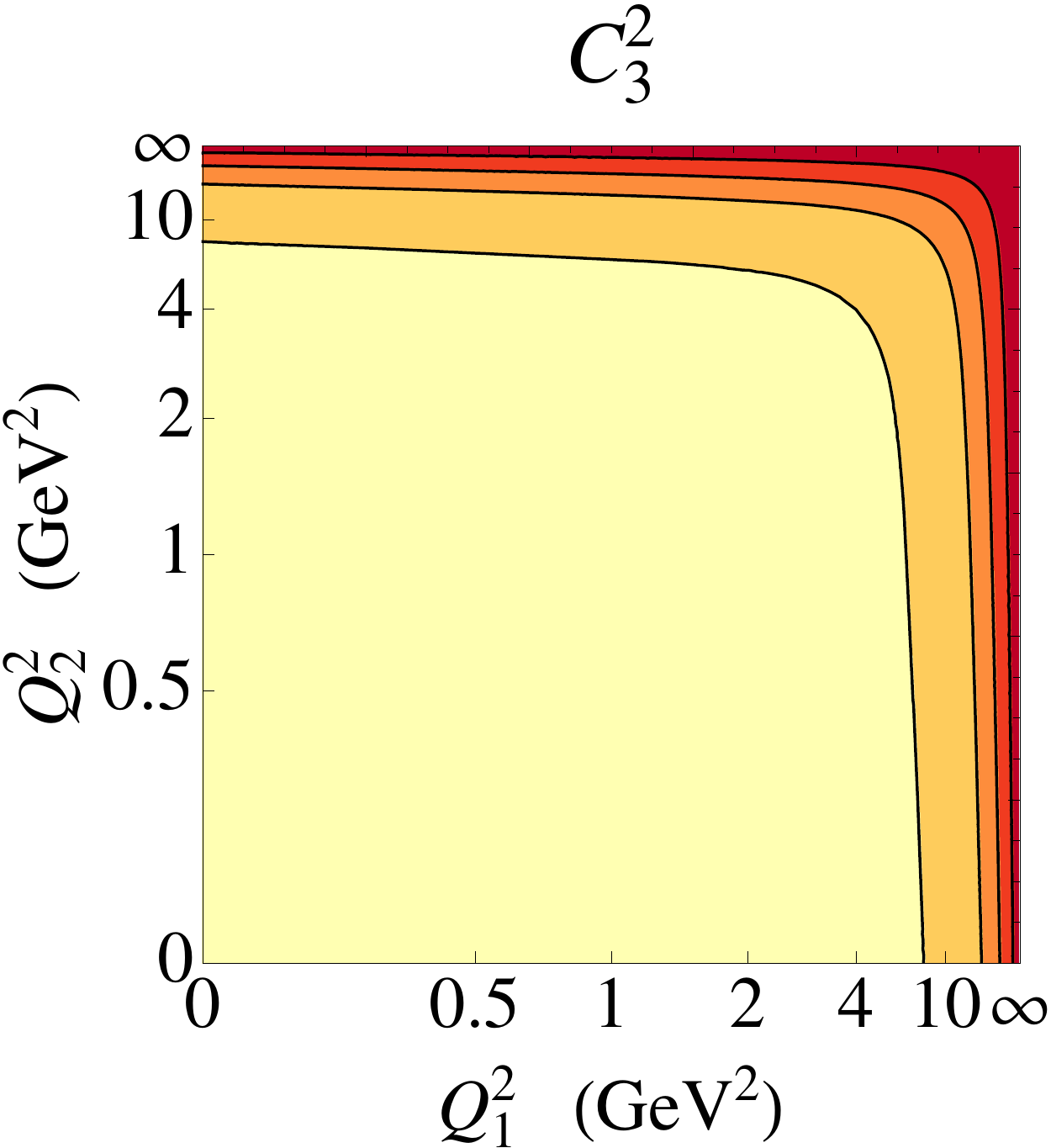}
  \caption{Convergence of the $C^{N}_{N+1}(Q_1^2,Q_2^2)$ sequence to the Regge model for the lowest elements. The first, second, third, and fourth contours, from light to dark red, 
  stand for the relative $-1,-5,-10$ and $-20\%$ deviations. Both axis have been scaled as $Q^2/(1+Q^2)$. \label{fig:cnn1regge}}
\end{figure}

\begin{figure}
\centering
  \includegraphics[width=0.325\textwidth]{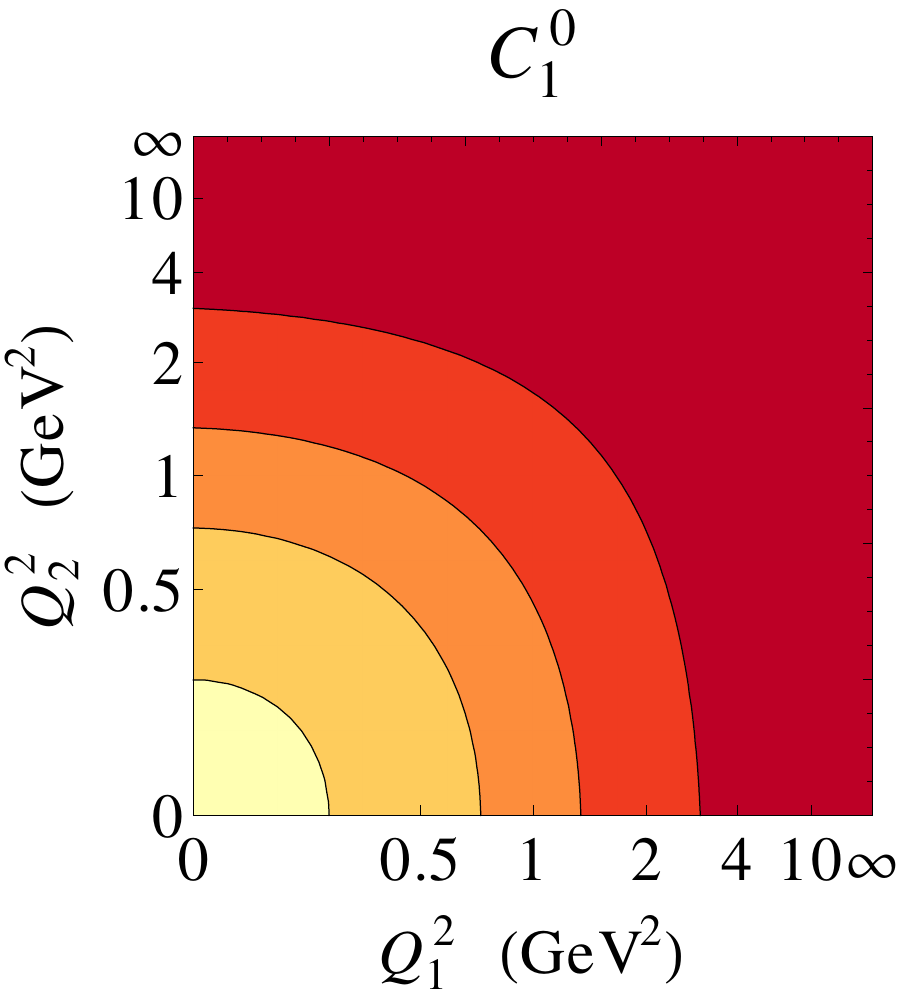}
  \includegraphics[width=0.325\textwidth]{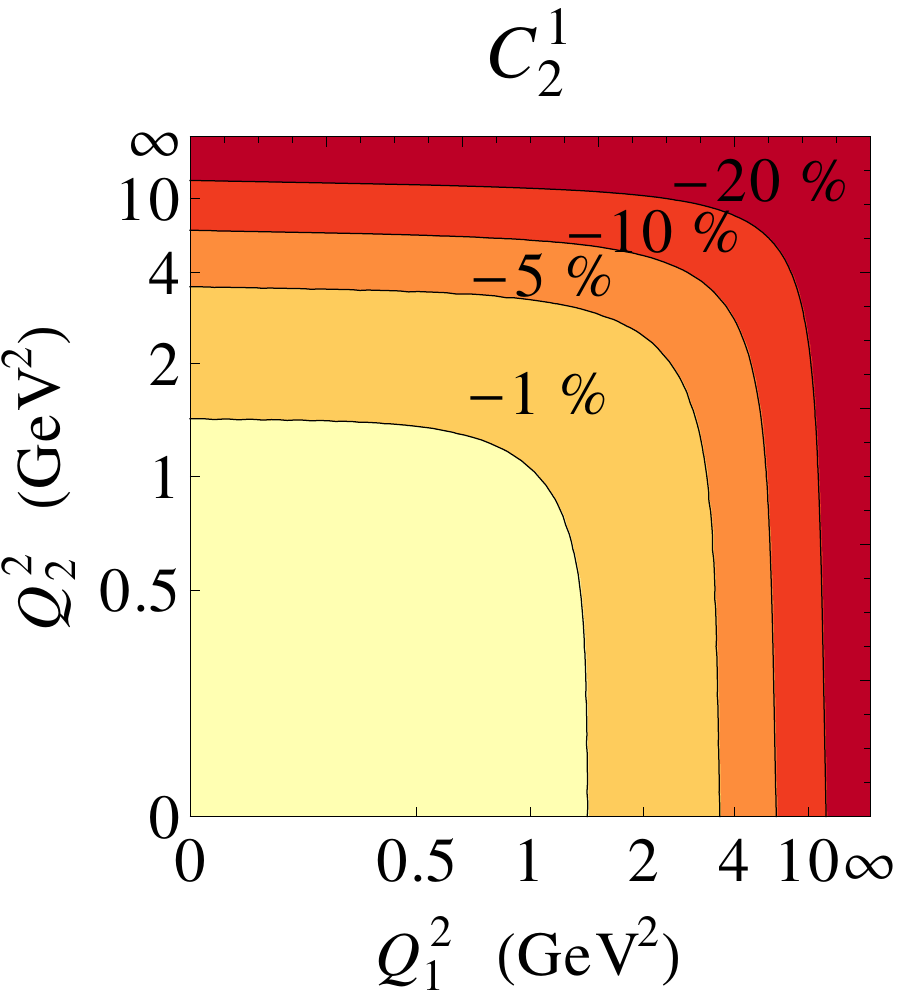}
  \includegraphics[width=0.325\textwidth]{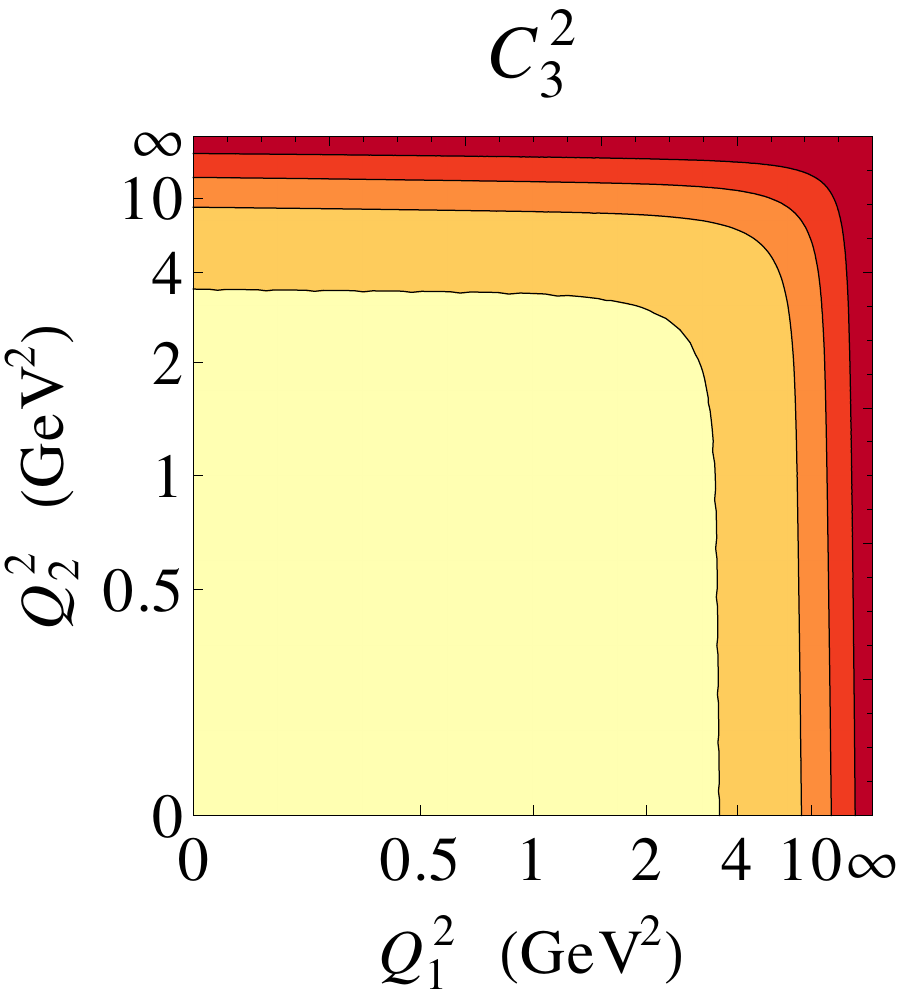}
  \caption{Analogous results to those in Fig.~\ref{fig:cnn1regge} but for the logarithmic model. \label{fig:cnn1log}}
\end{figure}

To illustrate the performance of Canterbury approximants, we take the example of two well-motivated models, the Regge model from Ref.~\cite{RuizArriola:2006ii} 
and a logarithmic model which naturally arises for flat distribution amplitudes~\cite{Radyushkin:2009zg}. These models have been employed in Ref.~\cite{RuizArriola:2006ii} and Refs.~\cite{Masjuan:2012wy,Radyushkin:2009zg} to study the TFF SL data and are selected here given their sophisticated analytic structure. The first one reads
\begin{equation}
\label{eq:ReggeDV}
F^{\textrm{Regge}}_{P\gamma^*\gamma^*}(Q_1^2,Q_2^2) = \frac{aF_{P\gamma\gamma}}{Q_1^2-Q_2^2}
                                         \frac{\left[ \psi^{(0)}\left(\frac{M^2+Q_1^2}{a}\right) -\psi^{(0)}\left(\frac{M^2+Q_2^2}{a}\right) \right]}{\psi^{(1)}\left(\frac{M^2}{a}\right)},
\end{equation}
where $\psi^{n}(z)=\partial_z^{n+1} \ln \Gamma(z)$ is the polygamma function, and $M$ and $a$ are the parameters for the Regge trajectory $M_n^2 = M^2 + na$. 
For our case of study, we take $M=0.8$~GeV and $a=1.3~\textrm{GeV}^2$~\cite{Masjuan:2012gc}. The lowest $C^0_1(Q_1^2,Q_2^2)$ approximant is given, for this model, as
\begin{equation}
C^0_1(Q_1^2,Q_2^2) = \frac{F_{P\gamma\gamma}}{ 1-\frac{(Q_1^2+Q_2^2)\psi^{(2)}}{2a\psi^{(1)}}  - \frac{Q_1^2Q_2^2}{2a^2}( \frac{\psi^{(3)}}{3\psi^{(1)}} - (\frac{\psi^{(2)}}{\psi^{(1)}})^2 )},
\end{equation}
where $\psi^{(n)}\equiv\psi^{(n)}(M^2/a)$. The second, logarithmic model, reads
\begin{equation}
\label{eq:appell}
F^{\textrm{log}}_{P\gamma^*\gamma^*}(Q_1^2,Q_2^2) = \frac{F_{P\gamma\gamma}M^2}{Q_1^2-Q_2^2}\ln\left( \frac{1+Q_1^2/M^2}{1+Q_2^2/M^2} \right).
\end{equation}
For our study, we take $M^2=0.6~\textrm{GeV}^2$ in analogy to the single-virtual case~\cite{Radyushkin:2009zg,Masjuan:2012wy}.
Its $C^0_1(Q_1^2,Q_2^2)$ approximant is then given as
\begin{equation}
C^0_1(Q_1^2,Q_2^2) = \frac{F_{P\gamma\gamma}}{1+\frac{Q_1^2+Q_2^2}{2M^2} + \frac{Q_1^2Q_2^2}{6M^4}}.
\end{equation}
To show the convergence, we use the $C^N_{N+1}$ sequence. The reason for this particular choice is not a mathematical but a physical one, namely, that the TFF should vanish 
as $Q_{1,2}^2\rightarrow\infty$. The results are shown in terms of the relative error
$C^N_{N+1}(Q_1^2,Q_2^2)/F^{\textrm{Regge,log}}_{P\gamma^*\gamma^*}(Q_1^2,Q_2^2)-1$ in Figs.~\ref{fig:cnn1regge} and \ref{fig:cnn1log} for the Regge and logarithmic model, respectively.
The observed convergence pattern is excellent and lies within the expectations of PAs results. 

We warn however that fixing the poles in advance to lie at the physical resonances (which holds strictly in the large-$N_c$ limit alone) results in a slower convergence pattern in analogy to PAs~\cite{Masjuan:2007ay}. Consequently, we do not advice this practice if only a small set of derivatives are known.

\begin{figure}
\centering
  \includegraphics[width=0.325\textwidth]{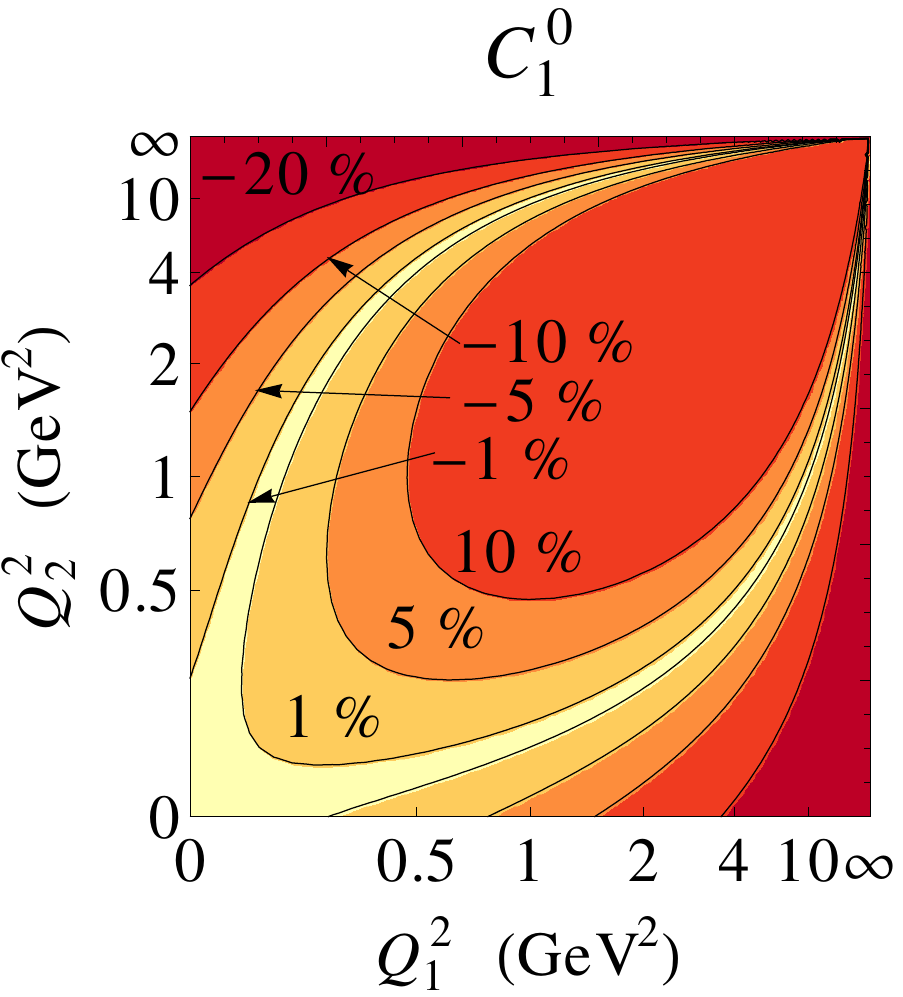}
  \includegraphics[width=0.325\textwidth]{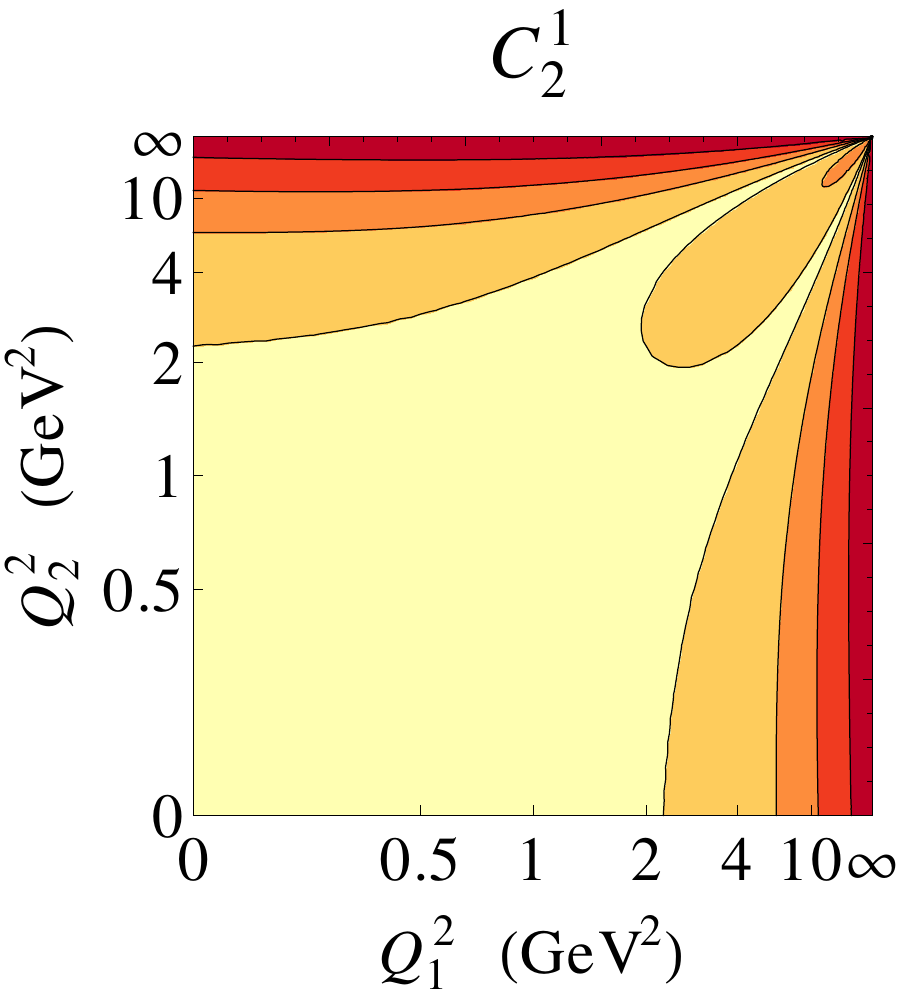}
  \includegraphics[width=0.325\textwidth]{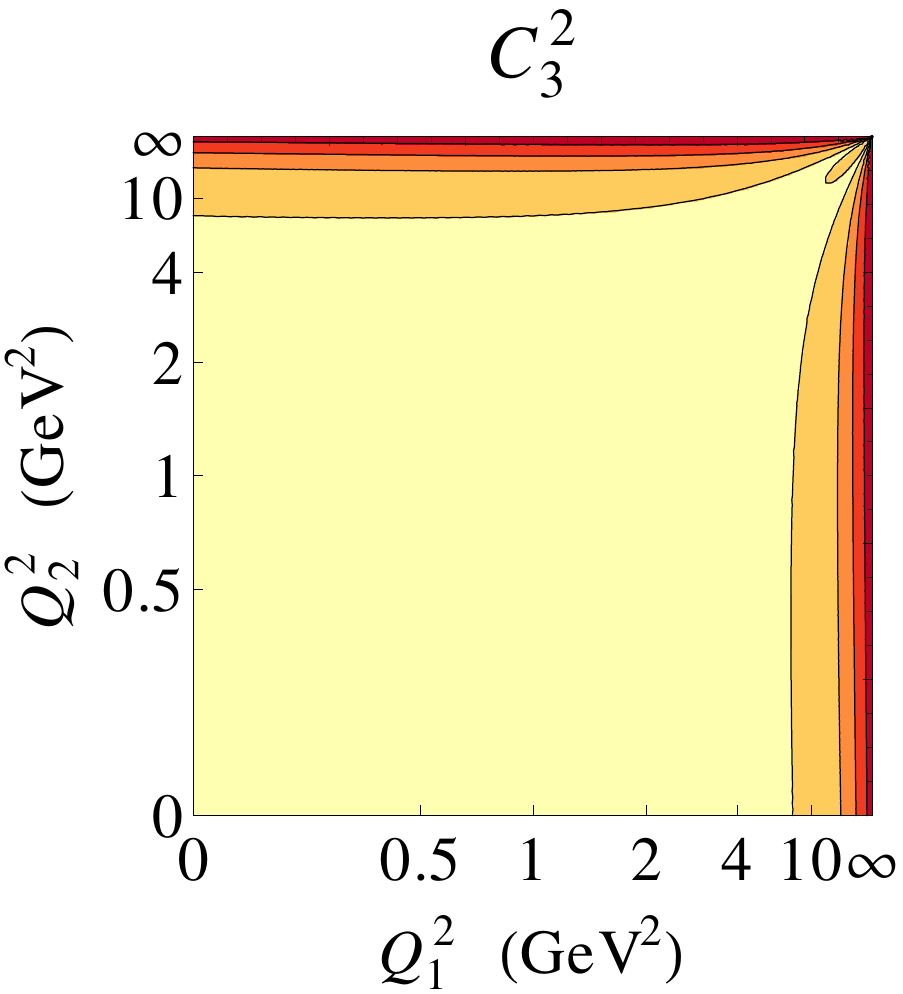}
  \caption{Convergence of the $C^N_{N+1}(Q_1^2,Q_2^2)$ sequence to the Regge model for the first elements as in Fig.~\ref{fig:cnn1regge} but now with the appropriate high-energy behavior imposed. The first, second, third, and fourth outer(inner) contours, from light to dark red, stand for the relative $\mp1,\mp5,\mp10$ and $\mp20\%$ deviations.  Both axis have been scaled as $Q^2/(1+Q^2)$. \label{fig:cnn1ope}}
\end{figure}

Finally, similar to Pad\'e approximants, the high-energy behavior may be constrained as well. Both of our models fulfill the OPE condition, meaning that, for 
$Q_1^2=Q_2^2\equiv Q^2\rightarrow \infty$, the TFF falls as $Q^{-2}$. As an example, for our
previous $C^0_1(Q_1^2,Q_2^2)$ case, this reduces to remove the $Q_1^2Q_2^2$ term in the denominator. The performance is greatly improved then for 
$Q_1^2=Q_2^2$ up to very large values, at least, for those approximants beyond $C^0_1(Q_1^2,Q_2^2)$, which is illustrated in Fig.~\ref{fig:cnn1ope}. 

In our case of study, we could only reconstruct the $C^0_1(Q_1^2,Q_2^2)$ approximant given the scarce information on the double-virtual regime. 
Particularly, we suggested in Ref.~\cite{Masjuan:2015lca} that, based on low- and high-energy constraints, 
the only involved double virtual parameter in this approximant, $a_{P;1,1}$, should lie within the  $(b_P^2,2b_P^2)$ range, which would provide then a conservative estimate 
for the BR$(P\to\bar{\ell}\ell)$ calculation.
In order to illustrate this statement, we display in Table~\ref{tab:modres} the range implied for these limiting cases (Fact and OPE columns) for our two given models together 
with the exact result of the models. 
\begin{table}[h]
\centering
\begin{tabular}{cccccccc} \hline
  & \multicolumn{4}{c}{Regge}  & \multicolumn{3}{c}{Log}   \\
  BR$(P\to\bar{\ell}\ell)$     & Fact & OPE & OPEc  &  \textbf{Exact}    & Fact & OPE & \textbf{Exact} \\ \hline
$\pi^0\to e^+e^-\times10^8$    & $6.218$ & $6.080$  & $6.266$ & $\boldsymbol{6.138}$  & $5.996$ & $5.869$ & $\boldsymbol{5.869}$ \\ 
$\eta\to e^+e^-\times10^9$    & $4.950$ & $5.064$  & $5.470$ & $\boldsymbol{5.012}$    & $4.614$ & $4.717$ & $\boldsymbol{4.626}$ \\ 
$\eta\to \mu^+\mu^-\times10^6$    & $4.844$ & $5.151$  & $4.688$ & $\boldsymbol{4.992}$     & $5.461$ & $5.889$ & $\boldsymbol{5.859}$ \\ 
$\eta'\to e^+e^-\times10^{10}$    & $1.825$ & $1.781$  & $1.867$ & $\boldsymbol{1.754}$     & $1.469$ & $1.437$ & $\boldsymbol{1.472}$ \\ 
$\eta'\to \mu^+\mu^-\times10^7$    & $1.518$ & $1.407$  & $0.922$ & $\boldsymbol{1.266}$    & $1.419$ & $1.405$ & $\boldsymbol{1.319}$ \\ \hline
\end{tabular}
\caption{Our Regge and logarithmic model results for the BR$(P\to\bar{\ell}\ell)$ (Exact column) together with their $C^0_1(Q_1^2,Q_2^2)$ approximant results. 
The Fact (OPE) column stands for our chosen values in the main text $a_{P;1,1}=b_P^2(2b_P^2)$, whereas the OPEc is a choice in which the OPE 
coefficient is imposed in the $C^0_1(Q_1^2,Q_2^2)$ reconstruction. See details in the text. \label{tab:modres}}
\end{table}
When calculating the models results, it is convenient to use alternative expressions equivalent to Eq.~\eqref{eq:ReggeDV} and Eq.~\eqref{eq:appell}. 
Particularly, for the Regge model we use
\begin{equation}
F_{P\gamma^*\gamma^*}^{\textrm{Regge}}(Q_1^2,Q_2^2)=\frac{F_{P\gamma\gamma}}{\psi^{(1)}(M^2/a)}\sum_{m=0}^{\infty} \frac{a^2}{(Q_1^2+(M^2+na))(Q_2^2+(M^2+na))}, 
\end{equation}
which allows to express the result in terms of a summation of individual terms which can be calculated analog to the factorization case. We find a nice convergence when summing up to  $10^{4}$ terms after comparison to the $10^{5}$ terms summation. For the logarithmic model, we employ the integral representation
\begin{equation}
F^{\textrm{log}}_{P\gamma^*\gamma^*}(Q_1^2,Q_2^2) = F_{P\gamma\gamma}M^2 \int_0^1 dx \frac{1}{xQ_1^2 + (1-x)Q_2^2 + M^2}.
\end{equation}
The loop integration can be performed using the same technique as in our OPE case; the resulting value is then integrated over $x$. From the results in Table~\ref{tab:modres}, the reader can verify that, for all the cases except the $\eta'$ ones, the model value lies within the suggested $a_{P;1,1}$ band. The differences in the $\eta'$ case are due to the ---thoroughly-discussed--- appearance of hadronic-induced imaginary parts and are within the error $\sim 20\%$ that was estimated in Table~\ref{tab:c01res}. Summarizing, our error estimations applied to our final results in Table~\ref{tab:mainres} are excellent for the models here investigated.

In addition, the reader may have realized that our $C^0_1(Q_1^2,Q_2^2)$ description can reproduce the BL and OPE $Q^2$-{\textit{behaviors}} but not their  associated coefficients at once. The latter would be possible with a  higher approximant~\cite{Masjuan:2015lca}, which is however difficult to reconstruct due to our ignorance on double-virtual parameters. Still, given the weight of the low energies in this process, we argued that this is of no relevance, since the low-energy parameters  play the main role. Our particular models chosen here, Eq.~\eqref{eq:ReggeDV} and Eq.~\eqref{eq:appell}, do not fulfill the BL $Q^2$-{\textit{behavior}} (they behave for large $Q^2$ as $Q^{-2}\ln Q^2$) but they do fulfill the OPE one. As a result, we can illustrate what would have been obtained if our $C^0_1(Q_1^2,Q_2^2)$ approximant fulfilling the OPE $Q^2${\textit{-behavior}} would have been constrained to reproduce the OPE coefficient as well ---this is, we trade the slope parameter, $b_P/m_P^2$, for the OPE coefficient. Unfortunately, for the logarithmic model this exercise is trivial, since $F_{P\gamma^*\gamma^*}(Q^2,Q^2) = M^2(M^2+Q^2)^{-1}$ and a single parameter can account for both, the low and high-energy behaviors. This is not the case for the Regge model, which result when constraining the OPE coefficient is given in the ``OPEc'' column in Table~\ref{tab:modres}. From this exercise we conclude that, given the weight of the  low energies, the most efficient strategy is to rely on the low-energy expansion and the correct high-energy behavior without imposing the high-energy coefficients. The OPEc always represents the worse scenario. This strategy is also supported from a mathematical point of view: whereas the Taylor expansion represents a convergent series with a finite radius of convergence, the high-energy expansion represents only an asymptotic one, for which the convergence of PAs is much slower~\cite{Masjuan:2007ay,Masjuan:2008fr}.
Let us finally remark that none of the single entries collected in Table~\ref{tab:modres} guide towards the exact result. It is only the range given by the two bounds (Fact $\div$ OPE) which tells where the result is.

\acknowledgments

Work supported by the Deutsche Forschungsgemeinschaft DFG through the Collaborative Research Center ``The Low-Energy Frontier of the Standard Model'' (SFB 1044).

\bibliographystyle{JHEP}
\bibliography{refs_pablo}

\end{document}